\documentclass{article}

\usepackage{PRIMEarxiv}
\usepackage{multirow,tabularx}
\usepackage[utf8]{inputenc} %
\usepackage[T1]{fontenc}    %
\usepackage{hyperref}       %
\usepackage{url}            %
\usepackage{booktabs}       %
\usepackage{amsfonts}       %
\usepackage{nicefrac}       %
\usepackage{microtype}      %
\usepackage{lipsum}
\usepackage{longtable}
\usepackage{fancyhdr}       %
\usepackage{graphicx}       %
\graphicspath{{figures/}}     %
\usepackage[font=small, labelfont=bf]{caption}
\usepackage{subcaption}
\usepackage{placeins}
\usepackage{soul,color}
\usepackage{amssymb}
\usepackage{amsmath}
\usepackage{pifont}
\newcommand{\cmark}{\ding{51}}
\newcommand{\xmark}{\ding{55}}
\newcommand\shorttitle{Crystal structure prediction using neural network potential and age-fitness Pareto genetic algorithm}
\newcommand\authors{Omee et al. ParetoCSP}

\title{Crystal structure prediction using neural network potential and age-fitness Pareto genetic algorithm
\thanks{\textit{\underline{Citation}}: 
\textbf{S.Omee et al. ParetoCSP.... DOI:000000/11111.}} 
}

\author{
 Sadman Sadeed Omee \\
 Department of Computer Science and Engineering\\
  University of South Carolina\\
  Columbia, SC 29201 \\
   \And
  Lai Wei\\
 Department of Computer Science and Engineering\\
  University of South Carolina\\
  Columbia, SC 29201 \\  
   \And
 Jianjun Hu *\\
 Department of Computer Science and Engineering\\
  University of South Carolina\\
  Columbia, SC 29201 \\
  \texttt{jianjunh@cse.sc.edu} \\
}

\begin{document}
\maketitle

\begin{abstract}

While crystal structure prediction (CSP) remains a longstanding challenge, we introduce ParetoCSP, a novel algorithm for CSP, which combines a multi-objective genetic algorithm (MOGA) with a neural network inter-atomic potential (IAP) model to find energetically optimal crystal structures given chemical compositions. We enhance the NSGA-III algorithm by incorporating the genotypic age as an independent optimization criterion and employ the M3GNet universal IAP to guide the GA search. Compared to GN-OA, a state-of-the-art neural potential based CSP algorithm, ParetoCSP demonstrated significantly better predictive capabilities, outperforming by a factor of $2.562$ across $55$ diverse benchmark structures, as evaluated by seven performance metrics. Trajectory analysis of the traversed structures of all algorithms shows that ParetoCSP generated more valid structures than other algorithms, which helped guide the GA to search more effectively for the optimal structures.

\end{abstract}


\section{Introduction}

Crystal structure prediction (CSP) is the problem of predicting the most energetically stable structure of a crystal given its chemical composition. Knowing the atomic structure is the most crucial aspect of comprehending crystalline materials. With the structural information of the material, advanced quantum-mechanical methods such as Density Functional Theory (DFT) can be utilized to calculate numerous physical characteristics of the crystal~\cite{oganov2006crystal}. As the physical and chemical characteristics of a crystal are dictated by the arrangement and composition of its atoms, CSP is critical to finding new materials that possess the needed properties such as high thermal conductivity, high compressing strength, high electrical conductivity, or low refractive index. CSP based computational materials discovery is significant and has the potential to revolutionize a range of industries, such as those involving electric vehicles, Li-batteries, building construction, energy storage, and quantum computing hardware~\cite{oganov2019structure,contactmap,searchmethods,brown2020artificial,kvashnin2019computational}. For this reason, CSP, along with  machine learning (ML)-based inverse design~\cite{noh2019inverse,brown2020artificial,kim2020inverse,dan2020generative,long2021constrained}, has emerged as one of the most potential methods for finding novel materials.

Although there have been notable advancements in the field of CSP, the scientific community has yet to solve this fundamental challenge that has persisted for decades. CSP presents a significant challenge due to the requirement to search through an extensive range of potential configurations to identify the most stable arrangement of atoms of a crystal in a high-dimensional space. The complexity of CSP stems from the combinatorial nature of the optimization challenge, where the number of potential configurations grows exponentially with the number of atoms present in the crystal~\cite{oganov2006crystal}. Additionally, the prediction of the most stable structure relies on several factors, including temperature, pressure, and chemical composition, further increasing the intricacy of the problem. Historically, the main method for determining crystal structures was through experimental X-ray diffraction (XRD)~\cite{oviedo2019fast}, which is time-consuming, expensive, and sometimes impossible, particularly for materials that are difficult to synthesize.

Computational approaches for CSP provide a faster and more affordable alternative than experimental methods. A typical strategy involves searching for the crystal's lowest energy atomic arrangement by optimizing its potential energy surface (PES) using different search algorithms. However, in some cases, simpler metrics such as the cohesive energy or the formation energy of the structures can be used instead~\cite{searchmethods}. The highly non-convex nature of the PES, which can contain a vast number of local minima, reduces the efficiency of the search algorithms. Moreover, finding the global minimum of a PES is categorized as an NP-hard problem~\cite{airss}. Most research on the CSP problem concentrates on \textit{ab initio} techniques, which involve exploring the atomic configuration space to locate the most stable structure based on the first-principles calculations of the free energy of possible structures~\cite{woodley2008crystal,uspex,calypso}. Although these methods are highly accurate, the scalability and the applicability of these ab initio algorithms for predicting crystal structures remain a challenge. These methods are severely constrained because they rely on expensive first-principles density functional theory (DFT) calculations~\cite{dft1,dft2} to determine the free energy of candidate structures. Furthermore, these methods are only applicable for predicting structures of comparatively small systems ($< 10 - 20$ atoms in the unit cell). Although there are inexpensive models available to estimate the free energy, they tend to have a poor correlation with reality, which can result in an inaccurate search~\cite{uspex}. For example, state-of-the-art (SOTA) graph neural networks (GNNs) have demonstrated the capability to accurately predict the formation energy of candidate structures~\cite{xie2018crystal, chen2019graph, schutt2018schnet, park2020developing, omee2022scalable, choudhary2021atomistic}, their performance on predicting non-stable or meta-stable structures is significantly lower as they are usually trained with stable crystals.

Several search algorithms have been applied to the CSP problem, including random sampling~\cite{airss}, simulated annealing~\cite{sim1,sim2,sim3}, meta-dynamics~\cite{md1,md2}, basin hopping~\cite{bh1,bh2}, minima hopping~\cite{mh1}, genetic algorithm (GA)~\cite{ge1,ge2,uspex,ge3}, particle swarm optimization (PSO)~\cite{calypso}, Bayesian optimization (BO)~\cite{bo1,gnoa}, and deep learning (DL)~\cite{dl1,dl2}. Among them, the USPEX algorithm, developed by Glass et al.~\cite{uspex}, is a prominent CSP algorithm based on evolutionary principles, using natural selection and reproduction to generate new crystal structures. It incorporates a combination of three operators- heredity, mutation, and permutation to explore the configuration space. To evaluate candidate structures, they use ab initio free energy calculation using tools like VASP~\cite{vasp} and SIESTA~\cite{siesta} which are highly accurate, but extremely time consuming. Another important CSP algorithm named CALYPSO was devised by Wang et al.~\cite{calypso}, which employs a PSO algorithm to explore the energy landscape of crystal structures and identify the lowest energy structures. To accomplish this, they developed a special strategy for removing comparable structures and applied symmetry-breaking restrictions to boost search effectiveness. Both USPEX and CALYPSO methods have been successfully applied to predicting the crystal structures of diverse materials, including those under high-pressure conditions, complex oxides, alloys, and etc. 
The random sampling-based CSP algorithms have also demonstrated their effectiveness. For example, AIRSS presented by Pickard et al.~\cite{airss}, describes a scheme that generates different random crystal structures for different type of crystals and conducts DFT calculations on them to determine the most stable one. Another genre of CSP methods are template-based methods~\cite{tem1,tem2,tem3} which involves finding an existing crystal structure as the template using some heuristic methods, or the ML method, etc, which has a similar chemical formula and then replacing some of its atoms with different elements. However, the accuracy of these models is constrained by the diversity and availability of the templates, as well as the complexity of the target compound. Inspired by the recent success of DL-based methods in protein structure prediction~\cite{protein1,protein2,protein3}, a DL-based algorithm, AlphaCrystal ~\cite{dl2} has been designed to predict the contact map of a target crystal and then reconstruct its structure via a GA. However, the effectiveness of this model is constrained because its performance relies on the accuracy of the predicted space group, lattice parameters, and distance matrices. Moreover, it ultimately depends on the optimization algorithm for reconstructing the final structure from the contact map as it is unable to provide end-to-end prediction like DeepMind's AlphaFold$2$~\cite{protein2}.

Compared to previous DFT-based CSP algorithms such as USPEX and CALYPSO, a major progress in CSP is to use machine-learning potential models to replace the costly first principle energy calculation. Cheng et al.~\cite{gnoa} developed a CSP framework named GN-OA, in which a graph neural network (GNN) model was first trained to predict the formation energy and then an optimization algorithm was then used to search for the crystal structure with the minimum formation energy, guided by the GNN energy model. They show that the BO search algorithm produces the best results among all optimization algorithms. However, predicting formation energy using GNNs has its drawback as its performance largely depends on the dataset it is trained on. A structure search trajectory analysis \cite{cspmetric} also showed that current BO and PSO in GN-OA tend to generate too many invalid structures, which deteriorates its performance. While both USPEX and CALYPSO have been combined with ML potentials for CSP before GN-OA, they were only applicable to small crystal systems such as Carbon structures, Sodium under pressure, and Boron clusters~\cite{uspexwithpotential,calyposwithpotential} due to the limitation of their ML potential models. 
Recently, significant progress has been achieved in ML potentials for crystals \cite{teanet,teanet45,choudhary2023unified,m3gnet,deng2023chgnet} that can work with multi- element crystals and larger crystals systems. This  will bring unprecedented opportunities and promise for modern CSP research and materials discovery. For example, recent advancement in deep neural network-based energy potential (M3GNet IAP) \cite{m3gnet} has shown its capability to cover $89$ elements of the periodic table while the CHGNet~\cite{deng2023chgnet} model was pretrained on the energies, forces, stresses, and magnetic moments from the Materials Project Trajectory Dataset, consisting of $\sim 1.5$ million unstable and stable inorganic structures. It is intriguing to explore how well modern CSP algorithms based on these ML potential can perform. Inspired by this progress, we propose the ParetoCSP algorithm for CSP, which combines the M3GNet potential with the age-fitness pareto genetic algorithms for efficient structure search. In this algorithm,  candidate structures in the GA population are compared based on both the genotypic age and the formation energy, predicted by a neural network potential such as M3GNet or CHGNet. Compared to previous GN-OAs, we showed that the significant global search capability of our ParetoCSP allows it to achieve much better prediction performance.

Our contribution in this paper can be summarized as follows:
\begin{itemize}
    \item We develop an efficient ParetoCSP for CSP, which combines an  updated multi-objective GA (NSGA-III) by the inclusion of the age fitness Pareto optimization criterion and a neural network potential (M3GNet IAP), utilized to correlate crystal structures to their final energy.
    \item Our systematic evaluations on $55$ benchmark crystals show that ParetoCSP outperforms GN-OA by a factor of $2.562$ in terms of prediction accuracy.
    \item We reinforce GN-OA by replacing its formation energy predictor MEGNet with the M3GNet IAP final energy model and show that it improves the default GN-OA by a factor of $1.5$ in terms of prediction accuracy. We further demonstrated the significant improvement in the search capability of ParetoCSP by showing that ParetoCSP outperforms the updated GN-OA by a factor of $1.71$ in terms of prediction accuracy.
    \item We provide quantitative analysis of the structures generated by ParetoCSP using seven performance metrics, and empirically show that ParetoCSP found better quality of structures for the test formulas than those by GN-OA.
    \item We perform a trajectory analysis of the generated structures by all evaluated CSP algorithms and show that ParetoCSP generates a great more valid solutions than the GN-OA algorithm, which may have contributed to ParetoCSP's better performance in predicting the crystal structures.
\end{itemize}

\section{Method}

\subsection{ParetoCSP: algorithm description}
The input of our algorithm (ParetoCSP) is the elemental composition of a crystal $\{c_i\}$, where $i$ is the index of an atom and $c_i$ is the element of the $i$-th atom in the unit cell. A periodic crystal
structure can be described by its lattice parameters ($L$) $a, b, c$ (representing the unit cell size), and $\alpha, \beta, \gamma$ (representing angles in the unit cell), the space group, and the atomic coordinates at unique Wyckoff positions. 

\begin{figure}[th]
  \centering
  \includegraphics[width=0.8\linewidth]{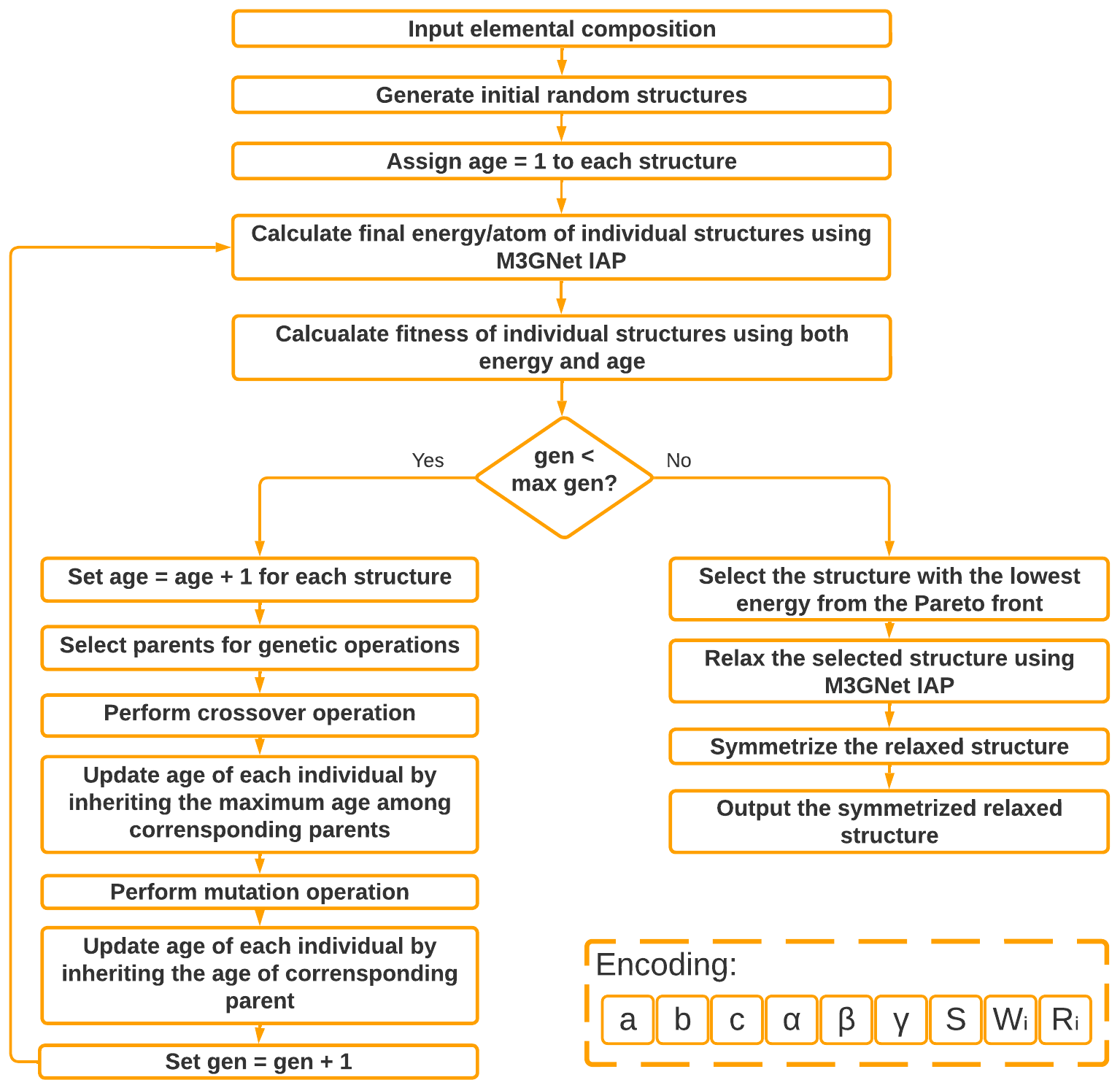}
  \caption{\textbf{The flowchart of ParetoCSP algorithm.}\\It starts by generating $n$ random crystals and assigning them an age of $1$, where $n$ denotes the population size. One complete generation then goes through the following steps: calculating energy of the structures and fitness, selecting parents, performing genetic operations, and updating the age. After a certain threshold of $\mathcal{G}$ generations, the lowest energy structure from the multi-dimensional Pareto front is chosen and further relaxed and symmetrized to obtain the final optimal structure. The genetic encoding is shown in the lower right corner of the flowchart. It contains lattice parameters $a$, $b$, $c$, $\alpha$, $\beta$, and $\gamma$, the space group $S$, the wyckoff position combination $W_i$, and the atomic coordinates $R_i$ of atom indexed by $i$.}
  \label{fig:flowchart}
\end{figure}

Our algorithm is based on the idea of the GN-OA algorithm~\cite{gnoa} with two major upgrades including the multi-objective GA search algorithm and the use of M3GNet potential for energy calculation. GN-OA has been proven from previous researches that incorporating symmetry constraint expedites CSP~\cite{gnoa,spai}. Similar to the GN-OA approach, our method also considers crystal structure prediction with symmetry constraints. We incorporate two additional structural features, namely crystal symmetry $S$ and the occupancy of Wyckoff position $W_i$ for each atom $i$. These features are selected from a collection of $229$ space groups and associated $1506$ Wyckoff positions~\cite{hahn1983international}. The method begins by selecting a symmetry $S$ from the range of $P2$ to $P230$, followed by generating lattice parameters $L$ within the chosen symmetry. Next, a combination of Wyckoff positions $\{W_i\}$ is selected to fulfill the specified number of atoms in the cell. The atomic coordinates $\{R_i\}$ are then determined based on the chosen Wyckoff positions $\{W_i\}$ and lattice parameters $L$. To generate crystal structures, we need to tune the $S$, $\{W_i\}$, $L$, and  $\{R_i\}$ variables. 

By selecting different combinations of $S$, ${W_i}$, $L$, and ${R_i}$, we can generate a comprehensive array of possible crystal structures for the given ${c_i}$. In theory, determining the energy of these various structures and selecting the one with the least energy should be the optimal crystal arrangement. However, exhaustively enumerating all these structures becomes practically infeasible due to the staggering number of potential combinations. To address this complexity, a more practical approach involves iteratively sampling candidate structures from the design space, under the assumption that one of the sampled structures will emerge as the most stable and optimal solution. Consequently, we adopt an optimization strategy to guide this search process towards identifying the structure with the lowest energy. In particular, we utilize an genetic algorithm, NSGA-III~\cite{nsga3-1,nsga3-2}, improved by incorporating AFPO~\cite{afpo} to enhance its performance and robustness.

First, we generate $n$ initial random structures. We then assign them an age of $1$ and convert them into crystal graphs. There are multiple approaches to encode crystals as graphs~\cite{isayev2017universal,xie2018crystal,chen2019graph,cheng2021geometric,matformer}. In short, we can consider each atom of the crystal as nodes of the graph, and interaction between them (e.g., bonds) can be encoded as edges. Interactions can be limited to certain cutoff range to define more realistic graphs. Each node and edge need to assigned feature vectors for the DNN to learn the specific property. After generating the initial structures, we predict their final energy/atom using the M3GNet universal IAP~\cite{m3gnet}. Next we calculate fitness considering both energy and age of the generated crystals (two independent dimension in the Pareto front). After that, we check whether the total number of generations are less than a certain threshold $\mathcal{G}$. If yes, we increase the age of all individuals by $1$. This follows the Pareto tournament selection, which selects the parents among the individual structures for the next generation. We usually set the tournament size to $2$ which selects half of the population as parents.

Next, we perform the genetic operations - crossover and mutation. After crossover, we update the age of each individual by inheriting the maximum age of corresponding parents. Similarly, after mutation, individual ages are updated by inheriting the age of their respective parents. These operations result in a new population of $n$ individuals for the next generation. The concept of age ensures a diverse population by containing both old and young individual, as well as effectively prevents from converging into local optima~\cite{afpo}. We then increase the generation number and repeat the whole process by calculating the final energy/atom of each structure until the generation number $\leq$ the threshold $\mathcal{G}$. After finishing $\mathcal{G}$. generations, we obtain a set of $\mathcal{F}$ non-dominated solutions on the Pareto front. We select the solution with the lowest final energy per atom as the optimal solution. We further relax the structure using the structure relaxation method of M3GNet IAP, which produces a more refined structure with lower final energy per atom. Finally, we perform a symmetrization operation to symmetrize the structure to output the final structure. Figure\ref{fig:flowchart} shows the flowchart of our ParetoCSP algorithm.

\subsection{AFPO: Age-fitness Pareto optimization}

One of the key requirements for a GA to achieve robust global search is to maintain the diversity of the population. Here, we employed the multi-objective genetic algorithm, AFPO by Schmidt and Lipson~\cite{afpo} to achieve this goal. The AFPO algorithm is inspired from the idea of age layered population structure (ALPS)~\cite{hu2005hierarchical,alpss}, which divides the evolving population into layers based on how long the genetic material has been present in the population so that competitions happen at different fitness levels, avoiding the occurrence of premature convergence. The \textit{age} of an individual is defined as how long the oldest part of its genotype has been present in the population~\cite{afpo-description}. Instead of partitioning the population into layers as done in the HFC algorithm \cite{hu2005hierarchical}, AFPO uses age as an explicit optimization criterion (an independent dimension in a multi-objective Pareto front). A solution is considered optimal if it has both higher fitness and lower age compared to other solutions. This enables the algorithm to maintain diversity in the population and avoid premature convergence to local optima, as well as to find better solutions at faster convergence speed~\cite{afpo}.

The AFPO algorithm starts by initializing a population of $N$ individuals randomly and assigned an age of one to all of them. The fitness of an individual is evaluated by calculating its performance for all objectives. The fitness values are then used to rank the individuals based on their Pareto dominance. The algorithm then updates and assigns the age for each individual. The age of an individual is increased by one with each generation. When crossover or mutation occurs, the individual's age is set to the maximum age of its parents. The algorithm uses a parameter called the tournament size $K$ which determines the number of individuals that compete for selection. Specifically, $K$ individuals are selected at random. It then forms the Pareto front among them, and eliminating any dominated individuals. After that, crossovers and mutations are applied to the parents to generate offspring. The objective function values for each offspring are evaluated and the updated ages are assigned to each offspring. The newly generated offspring replace some of the older individuals in the population based on their age and fitness values. To avoid premature convergence towards sub-optimal solutions, a few new random individuals are added to the population in each generation to maintain diversity. The algorithm continues to iterate through the above steps until a stopping criterion is met, such as a maximum number of generations or a desired level of convergence. For more details, the readers are referred to the reference~\cite{afpo-description}.

\subsection{NSGA-III: multi-objective GA}
We use the NSGA-III~\cite{nsga3-1} algorithm to implement the age-fitness based genetic algorithm AFPO. NSGA-II is an improved version of the popular multi-objective evolutionary algorithm NSGA-II~\cite{nsga2}. Here we describe the NSGA-III framework as defined in reference~\cite{nsga3-1,nsga3-2}. The NSGA-III algorithm begins with defining a group of reference points. To create an offspring population $Q_i$ at generation $i$, the current parent population $P_i$ undergoes genetic operations. The resulting population, $P_i \cup Q_i$ is then sorted based on their nondomination levels ($F_1, F_2,$ and so on). The algorithm saves all members up to the last fully accommodated level, $F_k$ (considering all solutions from level ($k + 1$) onward are rejected) in a set called $\delta_i$. The individuals from $\delta_i\setminus F_k$ have already been chosen for the next set of candidates, while the remaining spots are filled by individuals from $F_k$. 

The selection process of NSGA-III is substantially altered from the approach used in NSGA-II. First, the objective values and reference points are normalized. Second, each member in $\delta_i$ is assigned a reference point based on its distance to the individual with a reference line formed by connecting the ideal point to the reference point. This method enables the determination of the number and positions of population members linked to each supplied reference point in $\delta\setminus F_k$. Next, Next, a niching technique is applied to pick individuals from $F_k$ who are underrepresented in $\delta_i\setminus F_k$ based on the results of the association process explained earlier. Reference points with the fewest number of associations in the $\delta\setminus F_k$ population are identified and corresponding points in the $F_k$ set are searched. These selected members from $F_k$ are then added to the population, one by one, until the required population size is achieved. Thus NSGA-III utilizes a different approach in contrast to NSGA-II to sustain diversity among population members by incorporating a set of well-distributed reference points that are provided initially and updated adaptively during the algorithm's execution~\cite{nsga3-2}. More implementation details can be found in the reference~\cite{nsga3-implementation}.

\subsection{M3GNet Inter-atomic Potential (IAP)}
Energy potential is one of the key components of modern CSP algorithms. Here we use
M3GNet~\cite{m3gnet}, which is a GNN based ML potential model that explicitly incorporates $3$-body interactions. This model combines the graph-based DL inter-atomic potential and the many-body features found in traditional IAPs with the flexible graph material representations. One notable distinction of M3GNet from previous material graph implementations is the inclusion of atom coordinates and the $3\times 3$ lattice matrix in crystals. These additions are essential for obtaining tensorial quantities like forces and stresses through the use of auto-differentiation.

In the M3GNet model, position-included graphs serve as inputs. Graph features include embedded atomic numbers of elements and pair bond distances. Like traditional GNNs, the node and the edge features are updated via the graph convolution operations. Our M3GNet potential was trained using both stable and unstable structures so that it can well capture the difference between these two. The precise and efficient relaxation of diverse crystal structures and the accurate energy prediction achieved by the M3GNet-based relaxation algorithm make it well-suited for large-scale and fast crystal structure prediction.

\subsection{Evaluation criteria}
Many earlier studies \cite{calypso, uspex, airss} have depended on manual structural examination and ab initio formation energy comparison to assess the performance of a Crystal Structure Prediction (CSP) algorithm. But these metrics do not address the situation that an algorithm may not find the exact solution for a crystal and it is not clear how much the generated structure is deviated from the ground truth structure. Usually previous works did not quantitatively report how good or bad a solution is. Also, if two algorithms fail to generate the exact crystal structure, these metrics do not describe which one is closer to finding the optimal solution. Recently, Wei et al.~\cite{cspmetric} proposed a set of performance metrics to measure CSP performance which alleviated this issue greatly. We used seven performance metrics from that work to measure the performance of our CSP algorithm and the baselines. The required data are the crystallographic information file (CIF) of both the the optimized and relaxed final structure generated by the CSP algorithm and its corresponding ground truth stable structure. Details about these performance metrics can be found in \cite{cspmetric}. They are shortly listed below:
\begin{enumerate}
    \item Energy distance (ED)
    \item Wyckoff position fraction coordinate root mean squared error distance (W$_{rmse}$)
    \item Wyckoff position fraction coordinate root mean absolute error (W$_{mae}$)
    \item Sinkhorn distance (SD)
    \item Chamfer distance (CD)
    \item Hausdorff distance (HD)
    \item Crystal fingerprint distance (FP)
\end{enumerate}

\section{Results}
Our objective is to demonstrate the effectiveness of ParetoCSP for crystal structure prediction by showing that the multi-objective AFPO GA enables a much more effective structure search method than the BO and PSO and that M3GNet IAP is a more powerful crystal energy predictor than the previous MEGNet model.

\subsection{Benchmark set description}
We selected a diverse set of $55$ stable structures available in the Materials Project database~\cite{materialsproject} with no more than $20$ atoms. Among them, $20$ are binary crystals, $20$ are ternary crystals, and $15$ are quarternary crystals. We chose the benchmark set based on multiple factors such as diversity of elements, diversity of space groups, special type of materials (e.g., perovskites), and usage in previous CSP literature etc. Supplemental Fig. S1a shows the diversity of the elements used in the benchmark set. Table~\ref{table:dataset} shows the detailed information about the $55$ chosen test crystals used in this work.

\renewcommand{\arraystretch}{1.4}
\begin{longtable}{l c c c c c c}
\caption{\textbf{Details of the \boldmath{$55$} benchmark crystals used in this work.}\\The first $20$ crystals are binary, second $20$ crystals are ternary, and last $15$ crystals are quarternary, and each of these types of crystals are separated by single horizontal lines. We can see that the ground truth final energies and the predicted final energies by M3GNet IAP are very close, demonstrating M3GNet's effectiveness as an energy predictor.}
\label{table:dataset}\\
\hline\hline
\textbf{Composition} & \multicolumn{1}{l}{\begin{tabular}[c]{@{}c@{}}\textbf{No. of}\\\textbf{atoms}\end{tabular}} & \textbf{Space group} & \multicolumn{1}{l}{\begin{tabular}[c]{@{}c@{}}\textbf{Formation energy}\\\textbf{(eV/atom)}\end{tabular}} & \multicolumn{1}{l}{\begin{tabular}[c]{@{}c@{}}\textbf{Final energy}\\\textbf{(eV/atom)}\end{tabular}} & \multicolumn{1}{l}{\begin{tabular}[c]{@{}c@{}}\textbf{M3GNet final energy}\\\textbf{(eV/atom)}\end{tabular}}\\ 
\hline\hline
TiCo & $2$ & $Pm-3m$ & $-0.401$ & $-7.9003$ & $-7.8986$\\
CrPd$_3$ & $4$ & $Pm-3m$ & $-0.074$ & $-6.3722$ & $-6.4341$\\
GaNi$_3$ & $4$ & $Pm-3m$ & $-0.291$ & $-5.3813$ & $-5.3806$\\
ZrSe$_2$ & $3$ & $P-3m1$ & $-1.581$ & $-6.5087$ & $-6.5077$\\
MnAl & $2$ & $Pm-3m$ & $-0.225$ & $-6.6784$ & $-6.7503$\\
NiS$_2$ & $6$ & $P6_3/mmc$ & $-0.4$ & $-4.7493$ & $-4.9189$\\
TiO$_2$ & $6$ & $P4_2/mnm$ & $-3.312$ & $-8.9369$ & $-8.9290$\\
NiCl & $4$ & $P6_3mc$ & $-0.362$ & $-3.8391$ & $-3.8899$\\
AlNi$_3$ & $4$ & $Pm-3m$ & $-0.426$ & $-5.7047$ & $-5.6909$\\
CuBr & $4$ & $P6_3/mmc$ & $-0.519$ & $-3.0777$ & $-3.0908$\\
VPt$_3$ & $8$ & $I4/mmm$ & $-0.443$ & $-7.2678$ & $-7.2638$\\
MnCo & $2$ & $Pm-3m$ & $-0.0259$ & $-7.6954$ & $-7.6963$\\
BN & $4$ & $P6_3/mmc$ & $-1.411$ & $-8.7853$ & $-8.7551$\\
GeMo$_3$ & $8$ & $Pm-3n$ & $-0.15$ & $-9.4398$ & $-9.3588$\\
Ca$_3$V & $8$ & $I4/mmm$ & $0.481$ & $-3.2942$ & $-3.1638$\\
Ga$_2$Te$_3$ & $20$ & $Cc$ & $-0.575$ & $-3.4181$ & $-3.4160$\\
CoAs$_2$ & $12$ & $P2_1/c$ & $-0.29$ & $-5.8013$ & $-5.7964$\\
Li$_2$Al & $12$ & $Cmcm$ & $-0.163$ & $-2.6841$ & $-2.6623$\\
VS & $4$ & $P6_3/mmc$ & $-0.797$ & $-7.1557$ & $-7.3701$\\
Ba$_2$Hg & $6$ & $I4/mmm$ & $-0.384$ & $-1.7645$ & $-1.7582$\\
\hline

SrTiO$_3$ & $5$ & $Pm-3m$ & $-3.552$ & $-8.0249$ & $-8.0168$\\
Al$_2$FeCo & $4$ & $P4/mmm$ & $-0.472$ & $-6.2398$ & $-6.2462$\\
GaBN$_2$ & $4$ & $P-4m2$ & $-0.675$ & $-7.0893$ & $-7.0918$\\
AcMnO$_3$ & $5$ & $Pm-3m$ & $-2.971$ & $-7.1651$ & $-7.8733$\\
PaTlO$_3$ & $5$ & $Pm-3m$ & $-2.995$ & $-8.1070$ & $-8.1012$\\
CdCuN & $3$ & $P-6m2$ & $0.249$ & $-4.0807$ & $-4.0228$\\
HoHSe & $3$ & $P-6m2$ & $-1.65$ & $-5.2538$ & $-5.2245$\\
Li$_2$ZnSi & $8$ & $P6_3/mmc$ & $0.0512$ & $-2.5923$ & $-2.6308$\\
Cd$_2$AgPt & $16$ & $Fm-3m$ & $-0.195$ & $-2.8829$ & $-2.8415$\\
AlCrFe$_2$ & $4$ & $P4/mmm$ & $-0.157$ & $-7.7417$ & $-7.6908$\\
ZnCdPt$_2$ & $4$ & $P4/mmm$ & $-0.444$ & $-4.0253$ & $-4.0164$\\
EuAlSi & $3$ & $P-6m2$ & $-0.475$ & $-6.9741$ & $-6.9345$\\
Sc$_3$TlC & $5$ & $Pm-3m$ & $-0.622$ & $-6.7381$ & $-6.7419$\\
GaSeCl & $12$ & $Pnnm$ & $-1.216$ & $-3.6174$ & $-3.6262$\\
CaAgN & $3$ & $P-6m2$ & $-0.278$ & $-4.5501$ & $-4.7050$\\
BaAlGe & $3$ & $P-6m2$ & $-0.476$ & $-3.9051$ & $-3.9051$\\
K$_2$PdS$_2$ & $10$ & $Immm$ & $-1.103$ & $-4.0349$ & $-4.0066$\\
KCrO$_2$ & $8$ & $P6_3/mmc$ & $-2.117$ & $-6.4452$ & $-6.4248$\\
TiZnCu$_2$ & $4$ & $P4/mmm$ & $-0.0774$ & $-4.4119$ & $-4.4876$\\
Ta$_2$N$_3$O & $6$ & $P6/mmm$ & $-0.723$ & $-9.3783$ & $-9.3848$\\
\hline

AgBiSeS & $4$ & $P4/mmm$ & $-0.404$ & $-3.7363$ & $-3.8289$\\
ZrTaNO & $4$ & $P-6m2$ & $-1.381$ & $-9.5450$ & $-9.5429$\\
MnAlCuPd & $4$ & $P4mm$ & $-0.3$ & $-5.8467$ & $-5.8774$\\
CsNaICl & $4$ & $P4/mmm$ & $-1.79$ & $-2.9280$ & $-2.9448$\\
DyThCN & $4$ & $P4/mmm$ & $-1.03$ & $-8.3316$ & $-8.3510$\\
Li$_2$MgCdP$_2$ & $6$ & $P-4m2$ & $-0.61$ & $-3.4699$ & $-3.4514$\\
SrWNO$_2$ & $5$ & $P4/mmm$ & $-1.88$ & $-7.2188$ & $-7.0886$\\
Sr$_2$BBrN$_2$ & $18$ & $R-3m$ & $-1.639$ & $-6.1437$ & $-6.1501$\\
ZrCuSiAs & $8$ & $P4/nmm$ & $-0.592$ & $-6.2924$ & $-6.2853$\\
NdNiSnH$_2$ & $10$ & $P6_3/mmc$ & $-0.599$ & $-4.7970$ & $-4.8101$\\
MnCoSnRh & $12$ & $F-43m$ & $-0.25$ & $-7.1676$ & $-7.1093$\\
Mg$_2$ZnB$_2$Ir$_5$ & $20$ & $P4/mbm$ & $-0.454$ & $-6.6614$ & $-6.6577$\\
AlCr$_4$GaC$_2$ & $8$ & $P-6m2$ & $-0.151$ & $-8.1314$ & $-8.1246$\\
Y$_3$Al$_3$NiGe$_2$ & $9$ & $P-62m$ & $-0.735$ & $-5.8214$ & $-5.8305$\\
Ba$_2$CeTaO$_6$ & $20$ & $C2/m$ & $-3.49$ & $-8.2048$ & $-8.2384$\\
\hline\hline
\end{longtable}

\FloatBarrier

\subsection{Performance analysis of ParetoCSP}\label{susec: paretocsp_results}
The default version of ParetoCSP uses M3GNet universal IAP as the final energy evaluator for the candidate structures to guide the AFPO-based GA to identify the most stable structure with the minimum energy. Our algorithm ParetoCSP predicted the exact structures for $17$ out $20$ binary crystals ($85\%$), $16$ out of $20$ ternary crystals ($80\%$), and $8$ out of $15$ quarternary crystals ($53.333\%$) (see Table~\ref{table:result_comparison}). Overall, ParetoCSP achieved an accuracy of $74.55\%$ among all $55$ test crystals for this research which is the highest among all evaluated algorithms ($\approx 1.71\times$ the next best algorithm). Details on comparison with other algorithms and energy methods are discussed in Subsection~\ref{subsec: comp2} and \ref{subsec: comp3}. The exact accuracy results for all algorithms are presented in Table~\ref{table:result_comparison}. All the structures were assigned \cmark(exact), or \xmark(non-exact) based on manual inspection which was predominantly done in the majority of the past literature \cite{gnoa,calypso}.

\begin{figure}[ht] 
    \centering
    \begin{minipage}[c]{0.24\textwidth}
        \centering
        \includegraphics[width=\textwidth]{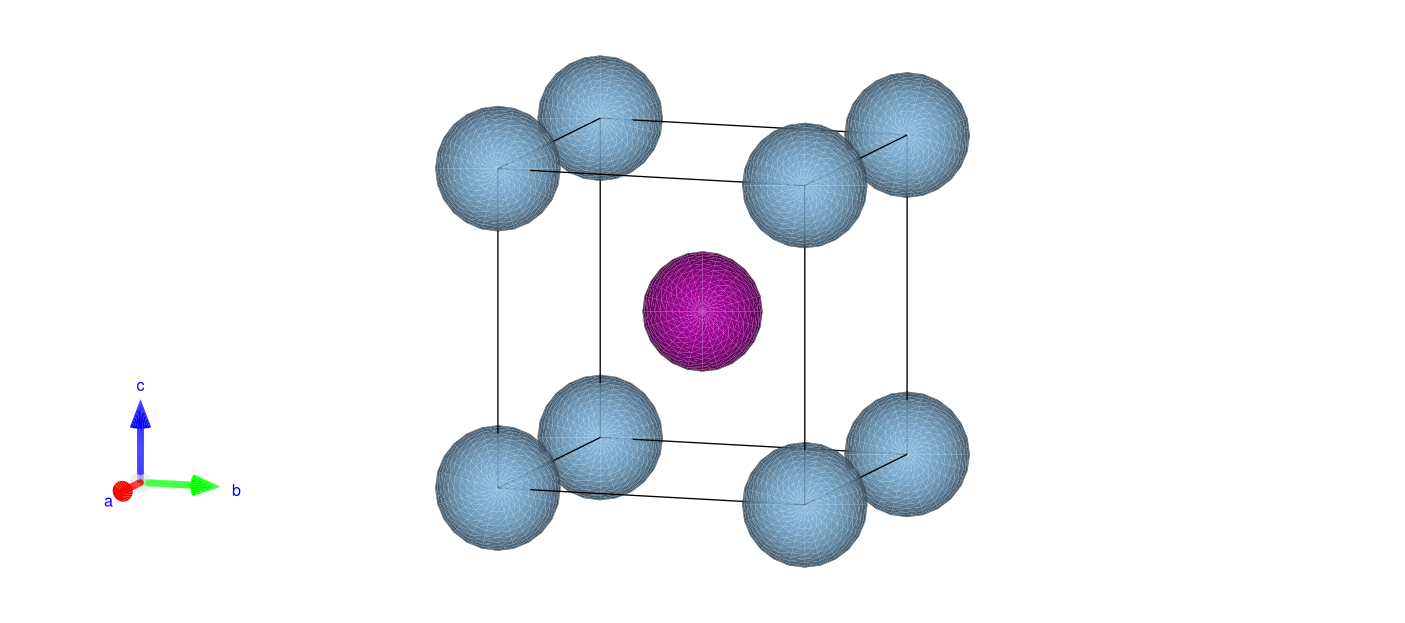}
        \subcaption{MnAl (ground truth)}
    \end{minipage}
    \begin{minipage}[c]{0.24\textwidth}
        \centering
        \includegraphics[width=\textwidth]{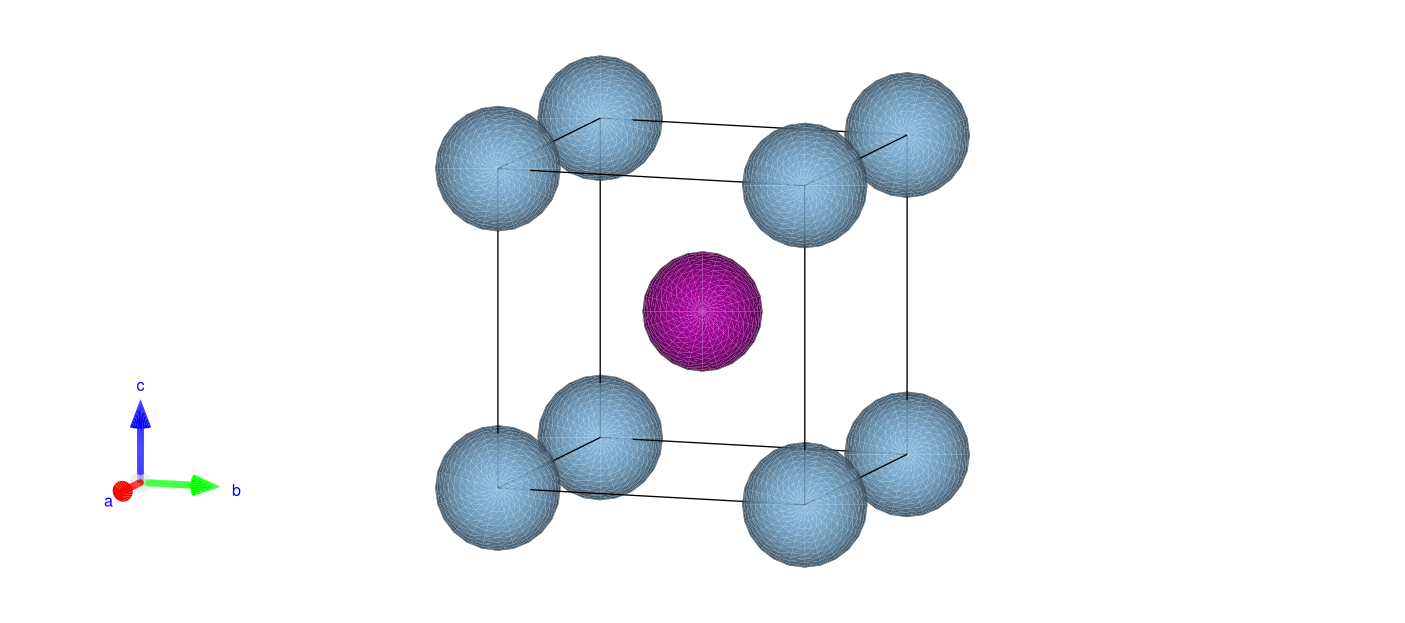}
        \subcaption{MnAl (predicted)}
    \end{minipage}
    \begin{minipage}[c]{0.24\textwidth}
        \centering
        \includegraphics[width=\textwidth]{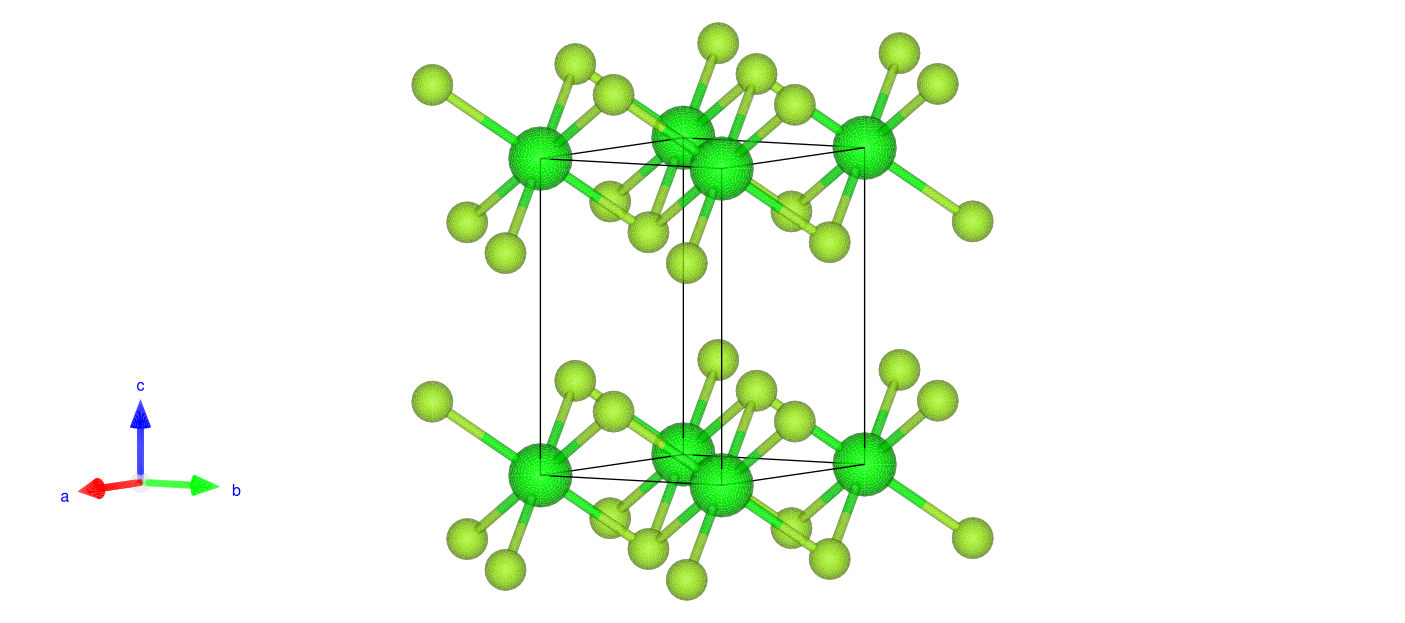}
        \subcaption{ZrSe$_2$ (ground truth)}
    \end{minipage}
    \begin{minipage}[c]{0.24\textwidth}
        \centering
        \includegraphics[width=\textwidth]{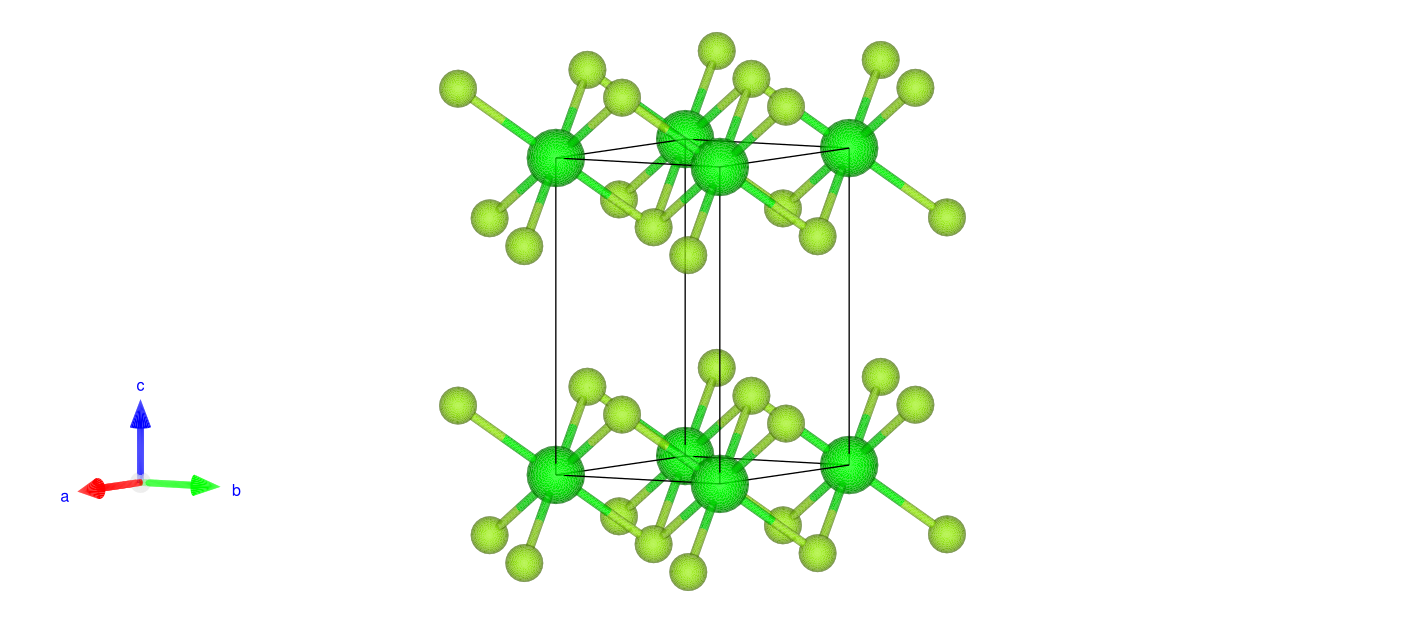}
        \subcaption{ZrSe$_2$ (predicted)}
    \end{minipage}\\

    \begin{minipage}[c]{0.24\textwidth}
        \centering
        \includegraphics[width=\textwidth]{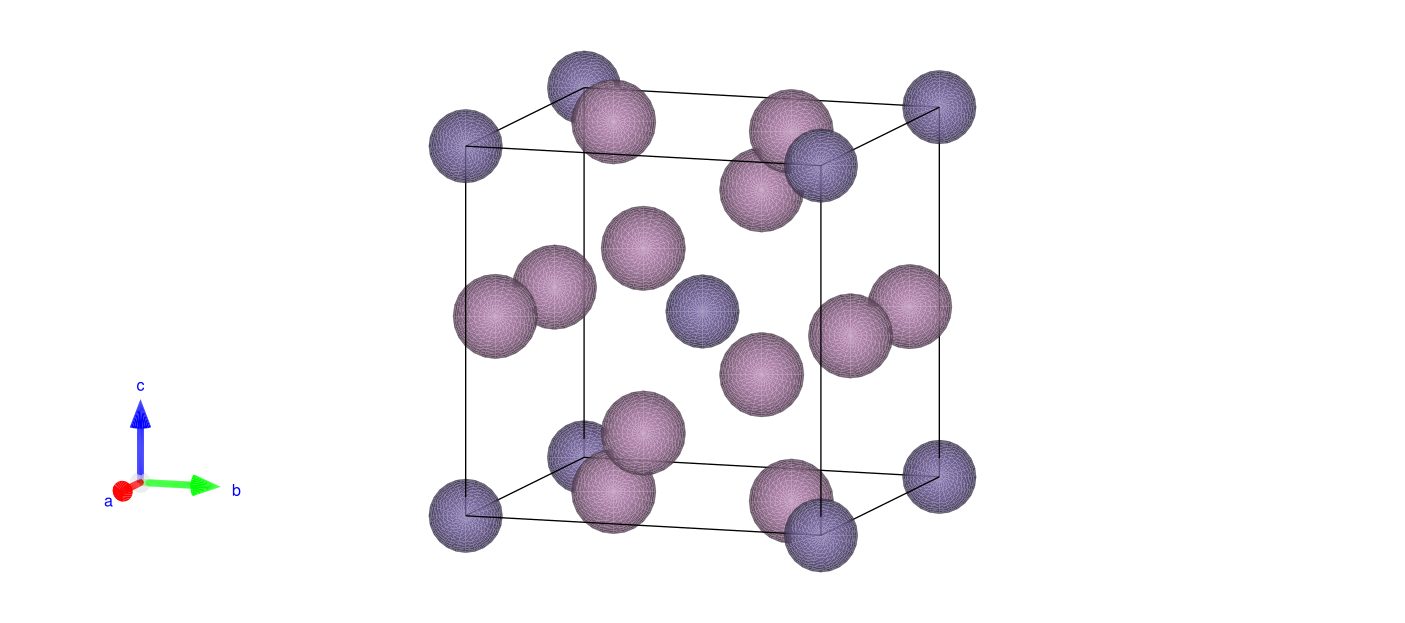}
        \subcaption{GeMo$_3$ (ground truth)}
    \end{minipage}
    \begin{minipage}[c]{0.24\textwidth}
        \centering
        \includegraphics[width=\textwidth]{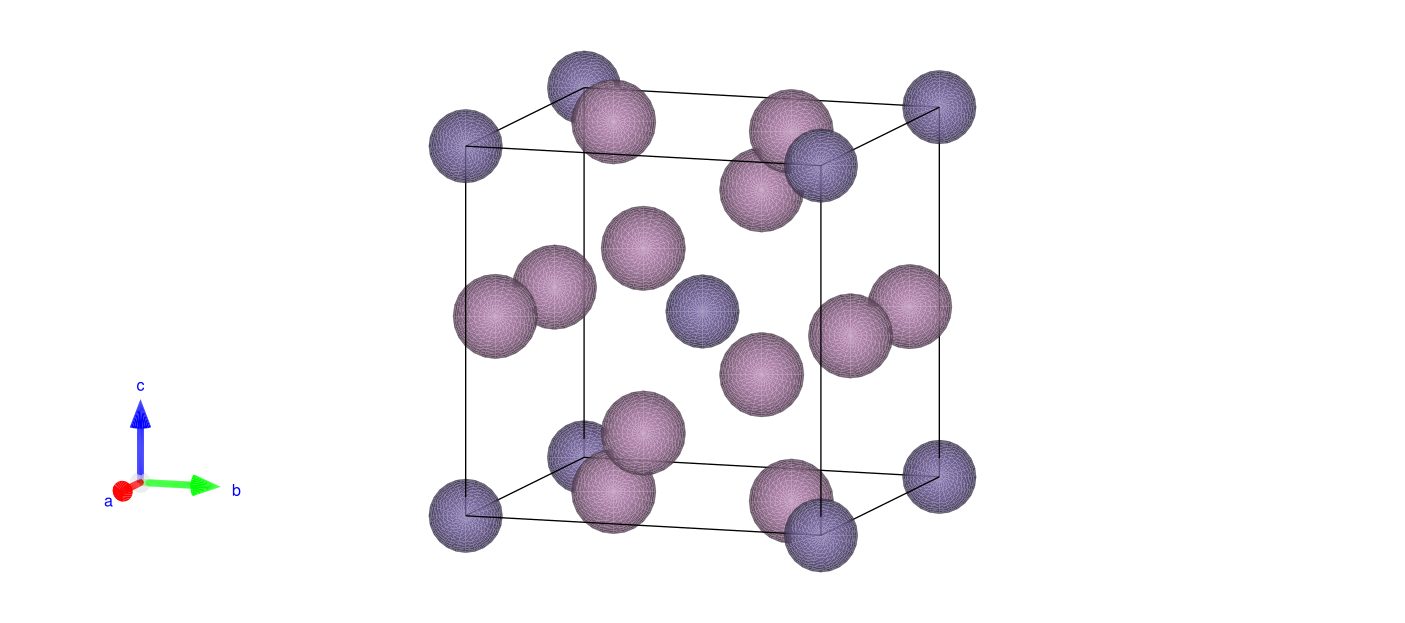}
        \subcaption{GeMo$_3$ (predicted)}
    \end{minipage}
    \begin{minipage}[c]{0.24\textwidth}
        \centering
        \includegraphics[width=\textwidth]{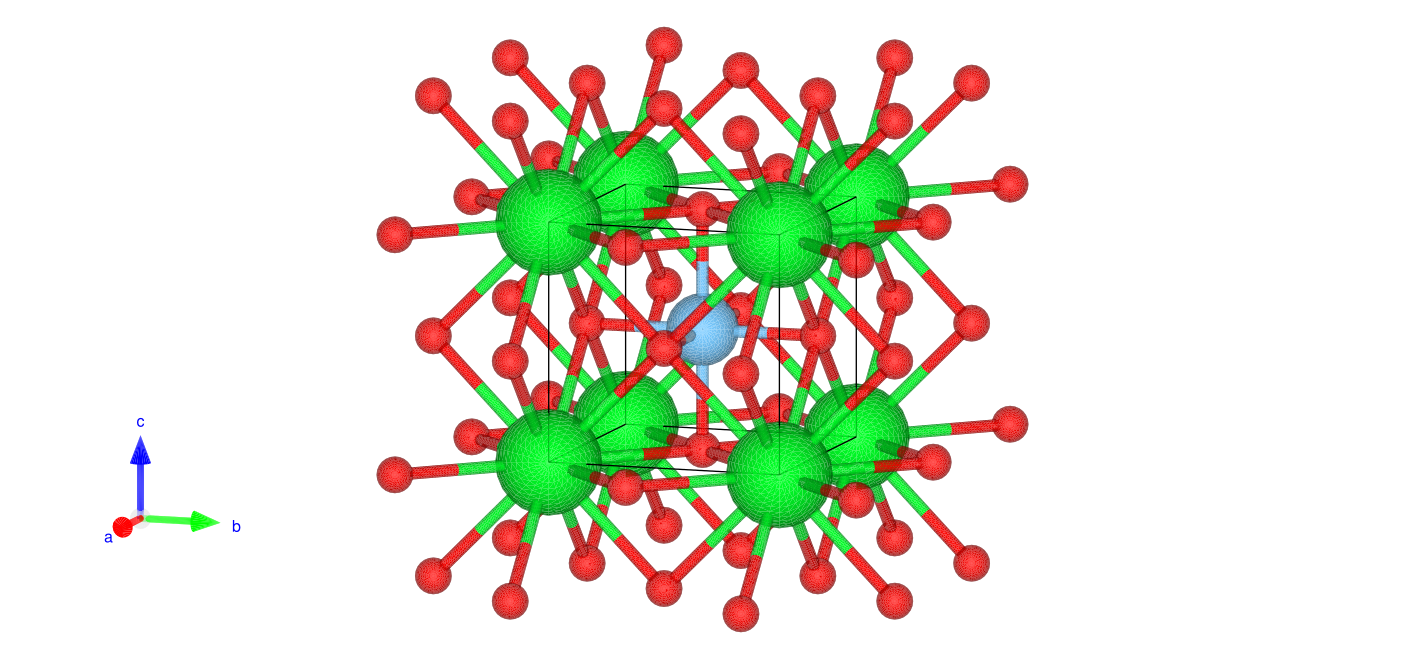}
        \subcaption{SrTiO$_3$ (ground truth)}
    \end{minipage}
    \begin{minipage}[c]{0.24\textwidth}
        \centering
        \includegraphics[width=\textwidth]{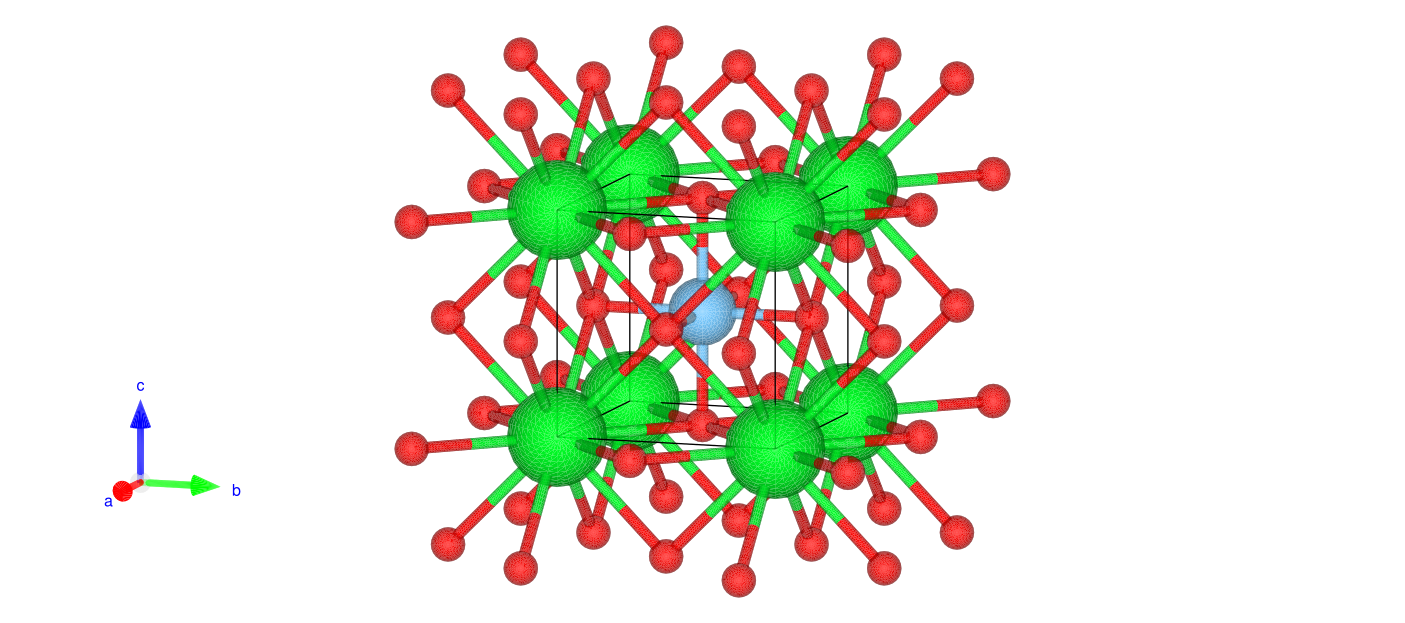}
        \subcaption{SrTiO$_3$ (predicted)}
    \end{minipage}\\

    \begin{minipage}[c]{0.24\textwidth}
        \centering
        \includegraphics[width=\textwidth]{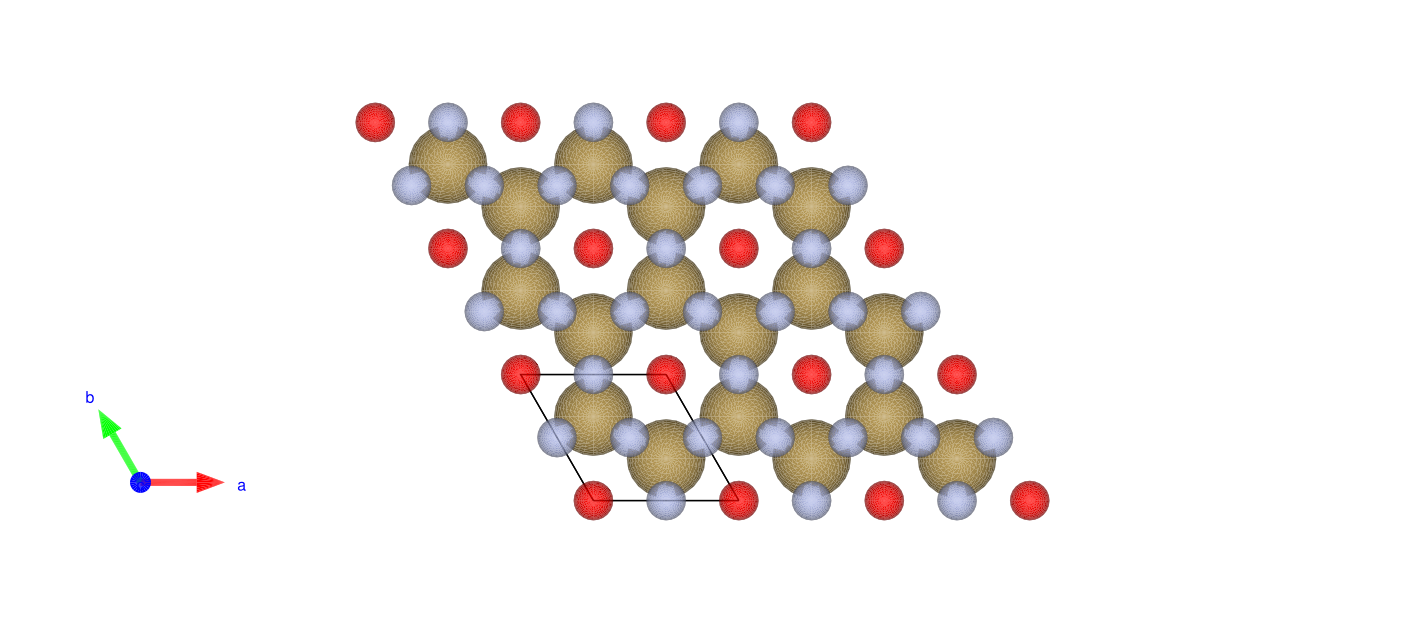}
        \subcaption{Ta$_2$N$_3$O (ground truth)}
    \end{minipage}
    \begin{minipage}[c]{0.24\textwidth}
        \centering
        \includegraphics[width=\textwidth]{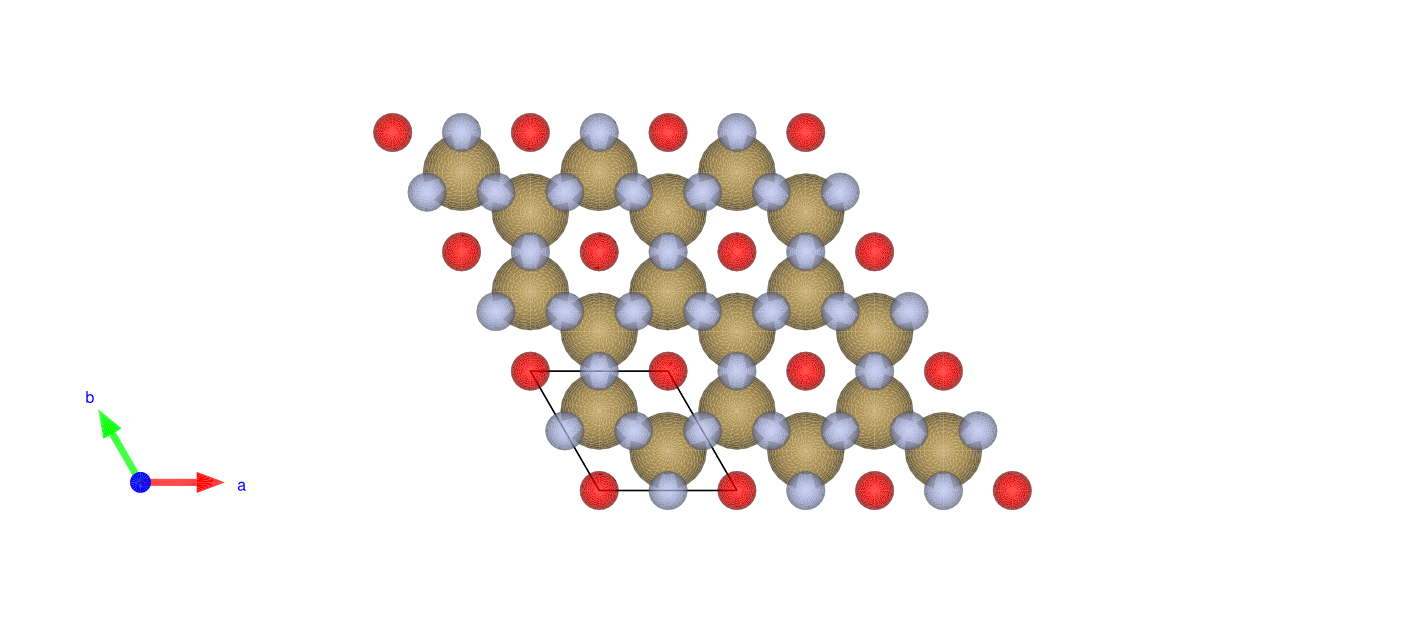}
        \subcaption{Ta$_2$N$_3$O (predicted)}
    \end{minipage}
    \begin{minipage}[c]{0.24\textwidth}
        \centering
        \includegraphics[width=\textwidth]{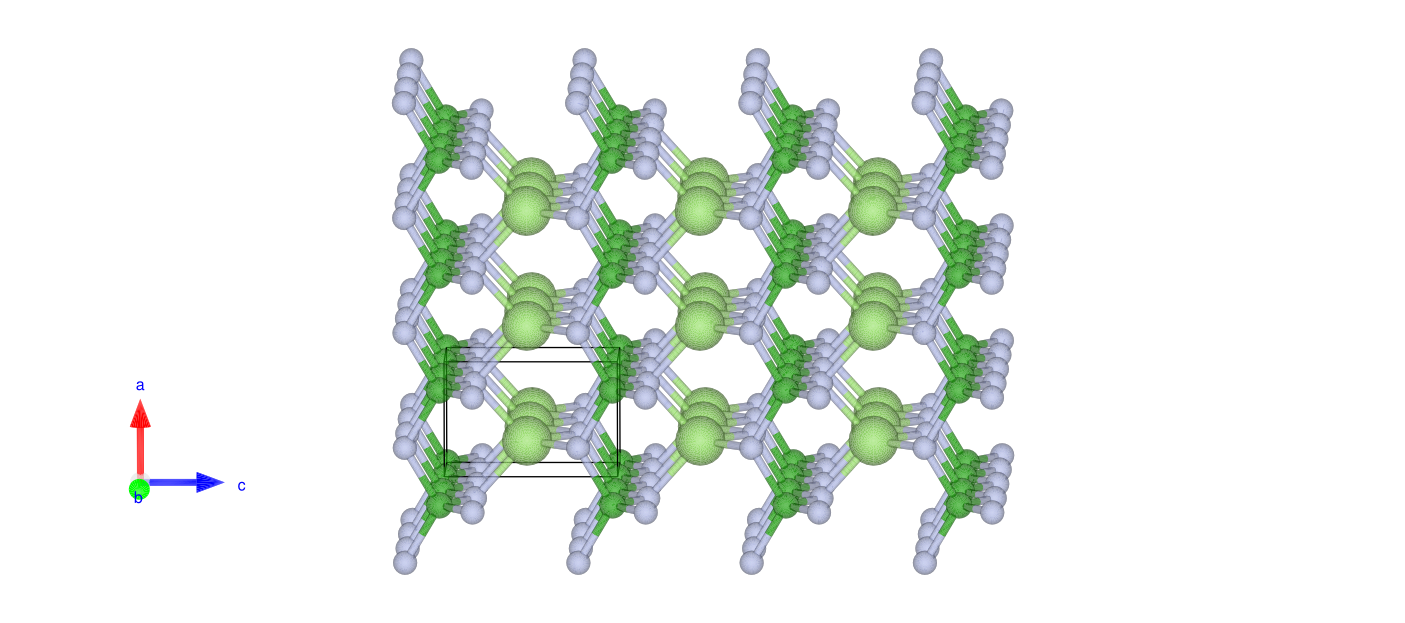}
        \subcaption{GaBN$_2$ (ground truth)}
    \end{minipage}
    \begin{minipage}[c]{0.24\textwidth}
        \centering
        \includegraphics[width=\textwidth]{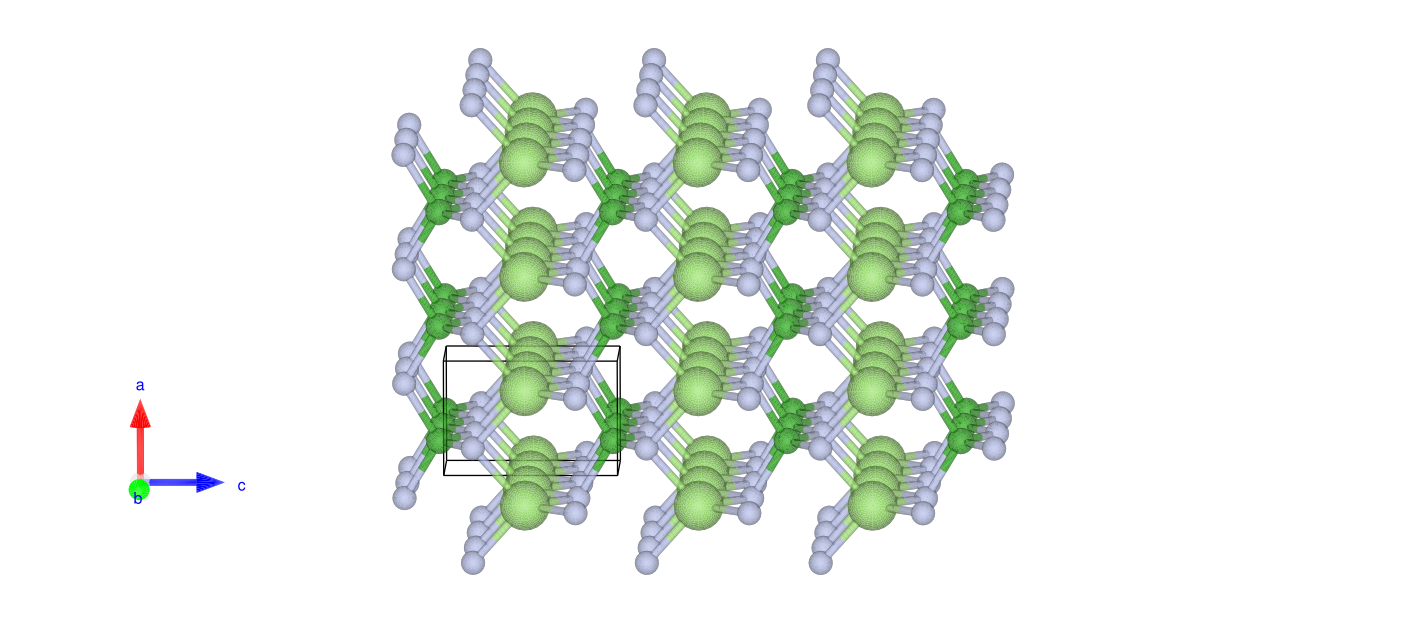}
        \subcaption{GaBN$_2$ (predicted)}
    \end{minipage}\\

    \begin{minipage}[c]{0.24\textwidth}
        \centering
        \includegraphics[width=\textwidth]{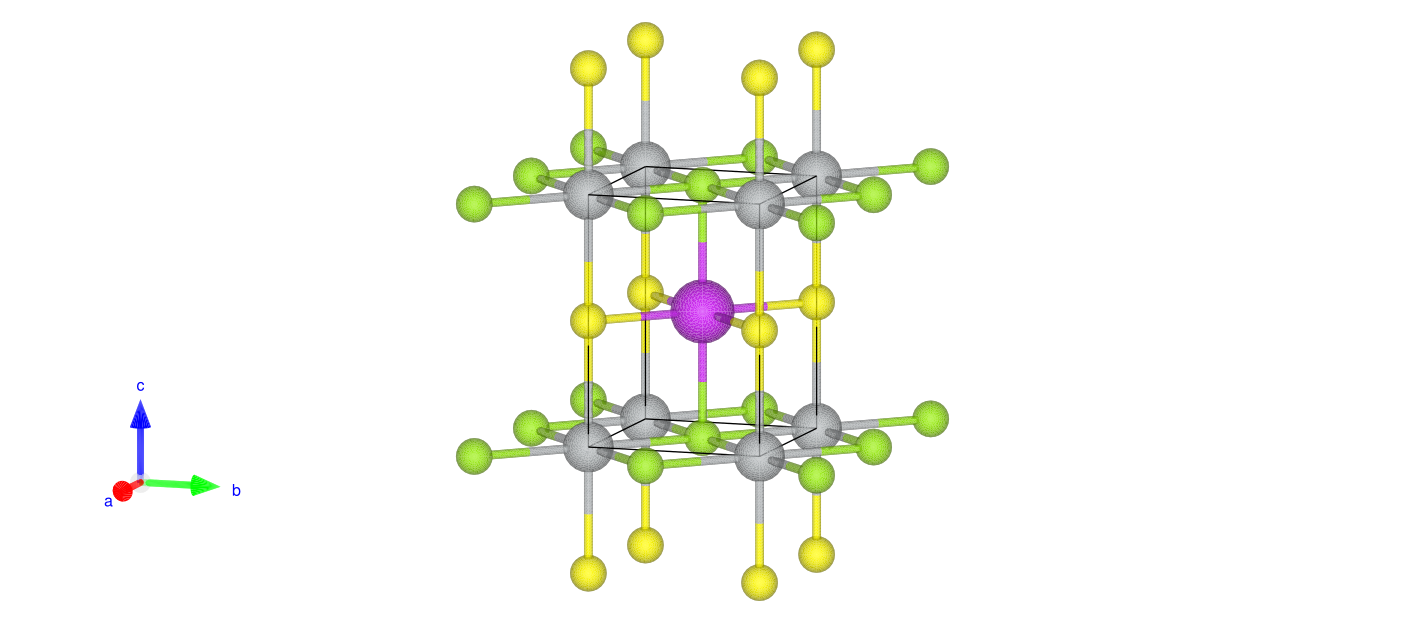}
        \subcaption{AgBiSeS (ground truth)}
    \end{minipage}
    \begin{minipage}[c]{0.24\textwidth}
        \centering
        \includegraphics[width=\textwidth]{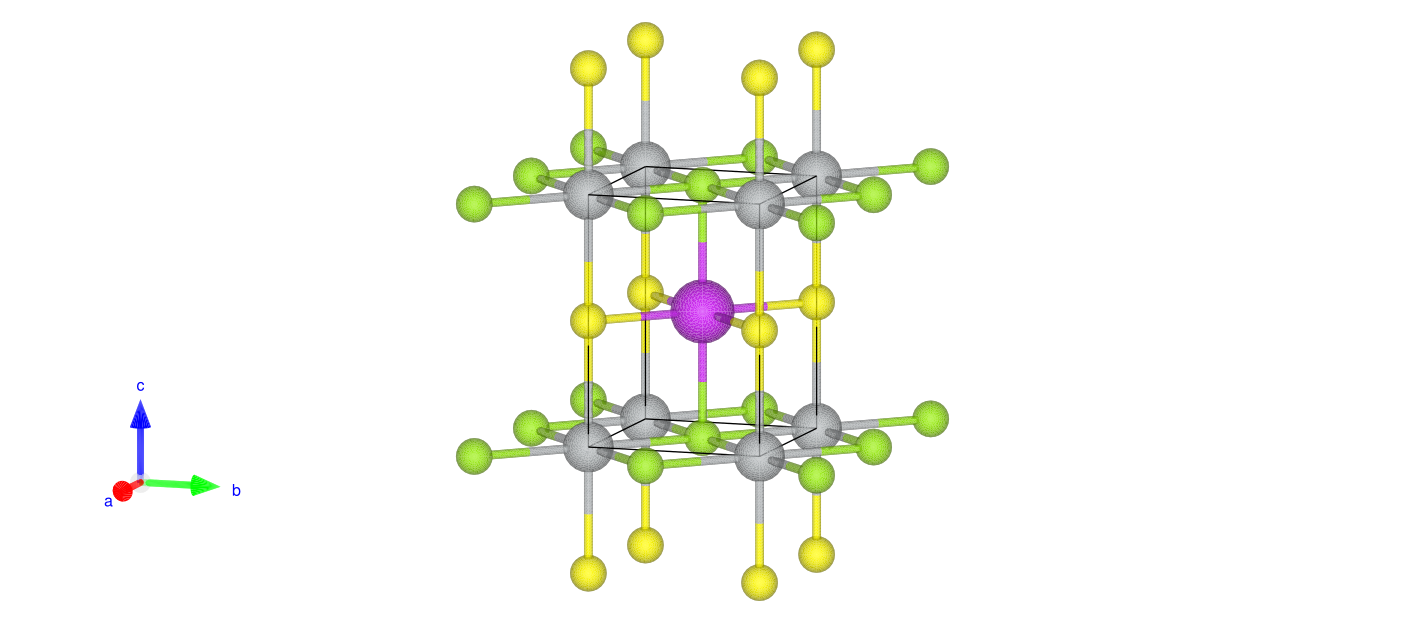}
        \subcaption{AgBiSeS (predicted)}
    \end{minipage}
    \begin{minipage}[c]{0.24\textwidth}
        \centering
        \includegraphics[width=\textwidth]{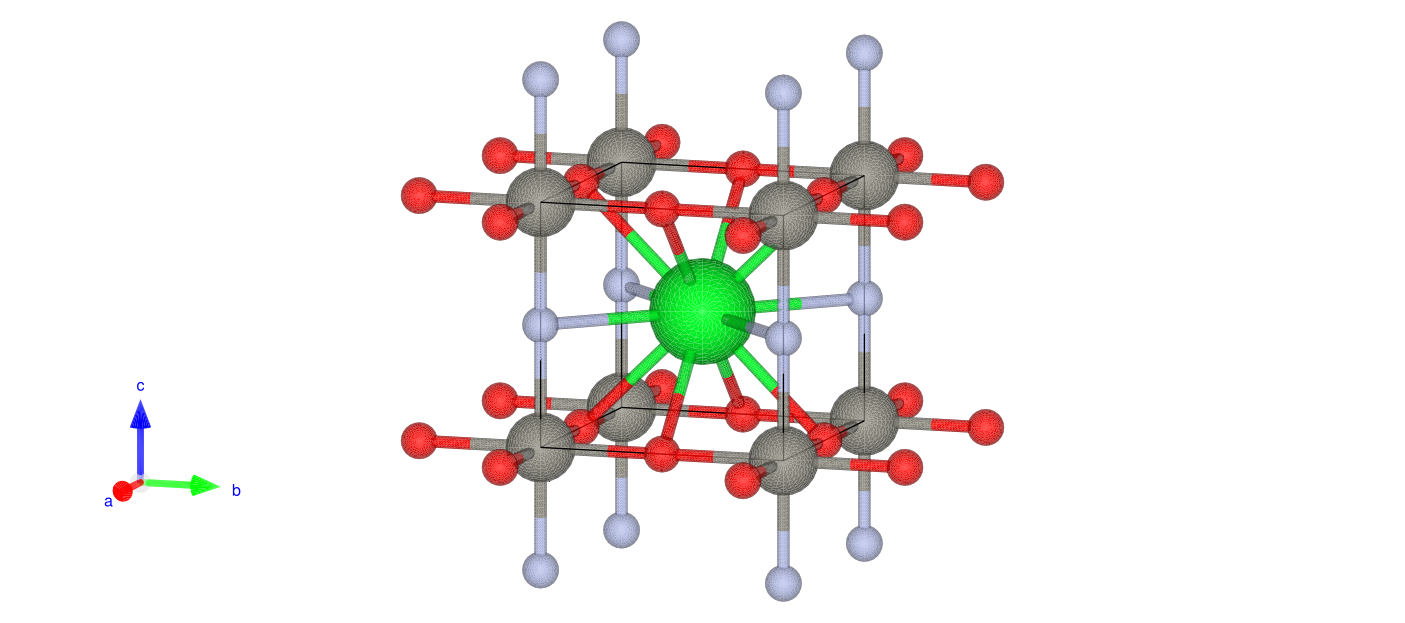}
        \subcaption{SrWNO$_2$ (ground truth)}
    \end{minipage}
    \begin{minipage}[c]{0.24\textwidth}
        \centering
        \includegraphics[width=\textwidth]{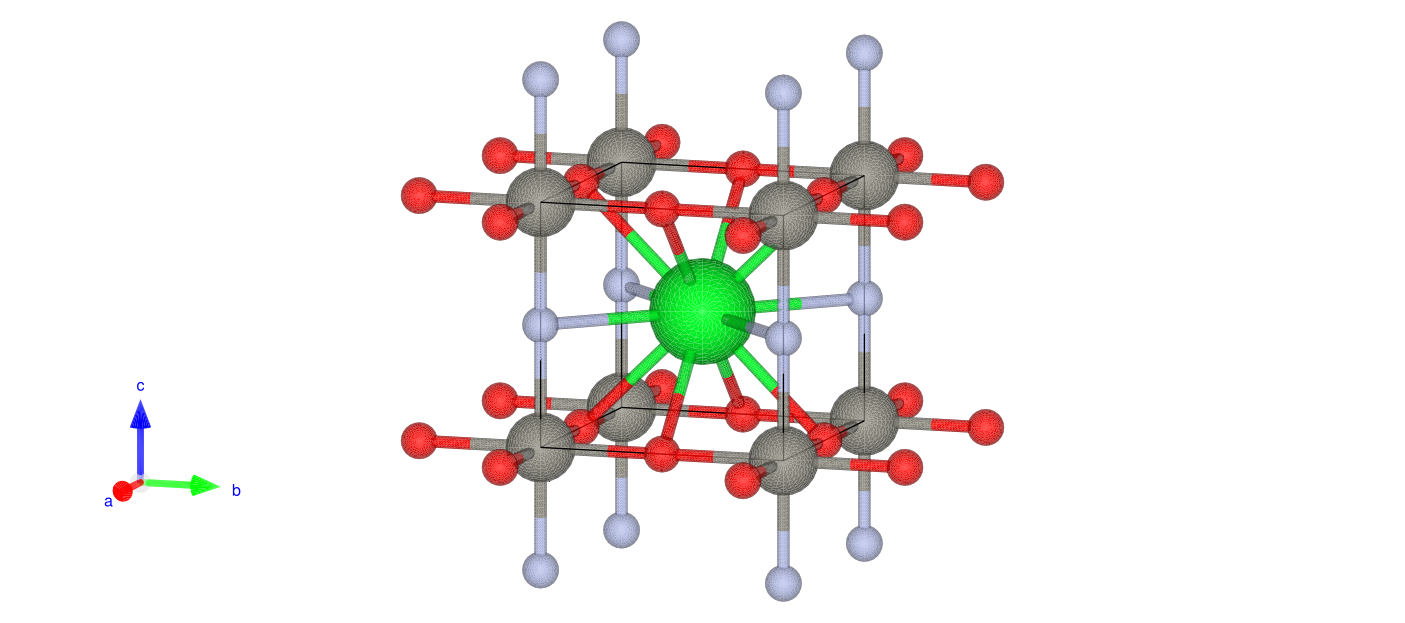}
        \subcaption{SrWNO$_2$ (predicted)}
    \end{minipage}\\

    \begin{minipage}[c]{0.24\textwidth}
        \centering
        \includegraphics[width=\textwidth]{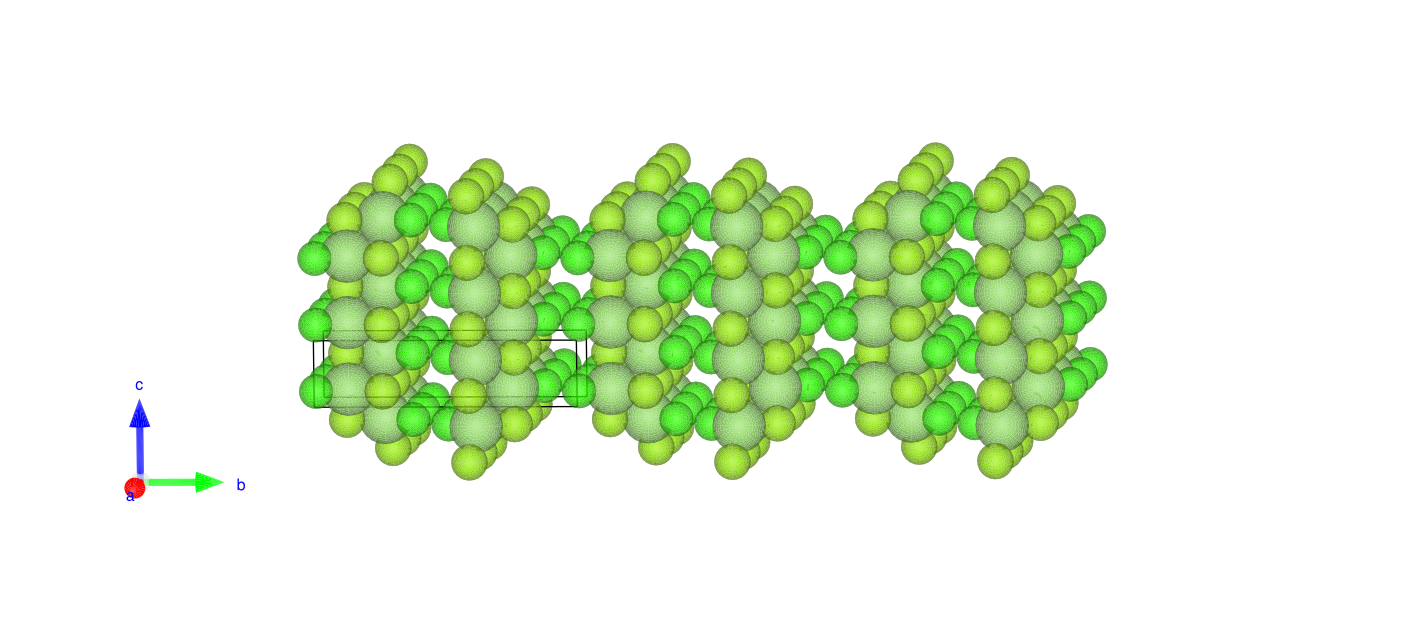}
        \subcaption{GaSeCl (ground truth)}
    \end{minipage}
    \begin{minipage}[c]{0.24\textwidth}
        \centering
        \includegraphics[width=\textwidth]{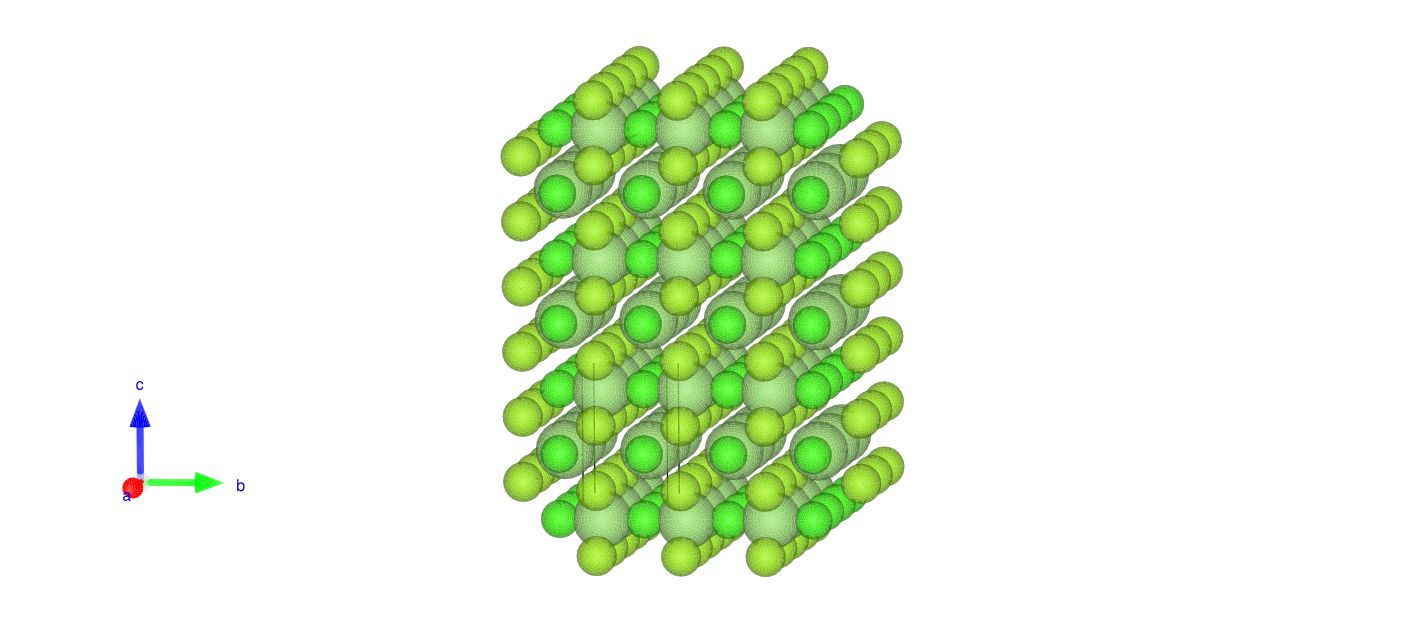}
        \subcaption{GaSeCl (predicted)}
    \end{minipage}
    \begin{minipage}[c]{0.24\textwidth}
        \centering
        \includegraphics[width=\textwidth]{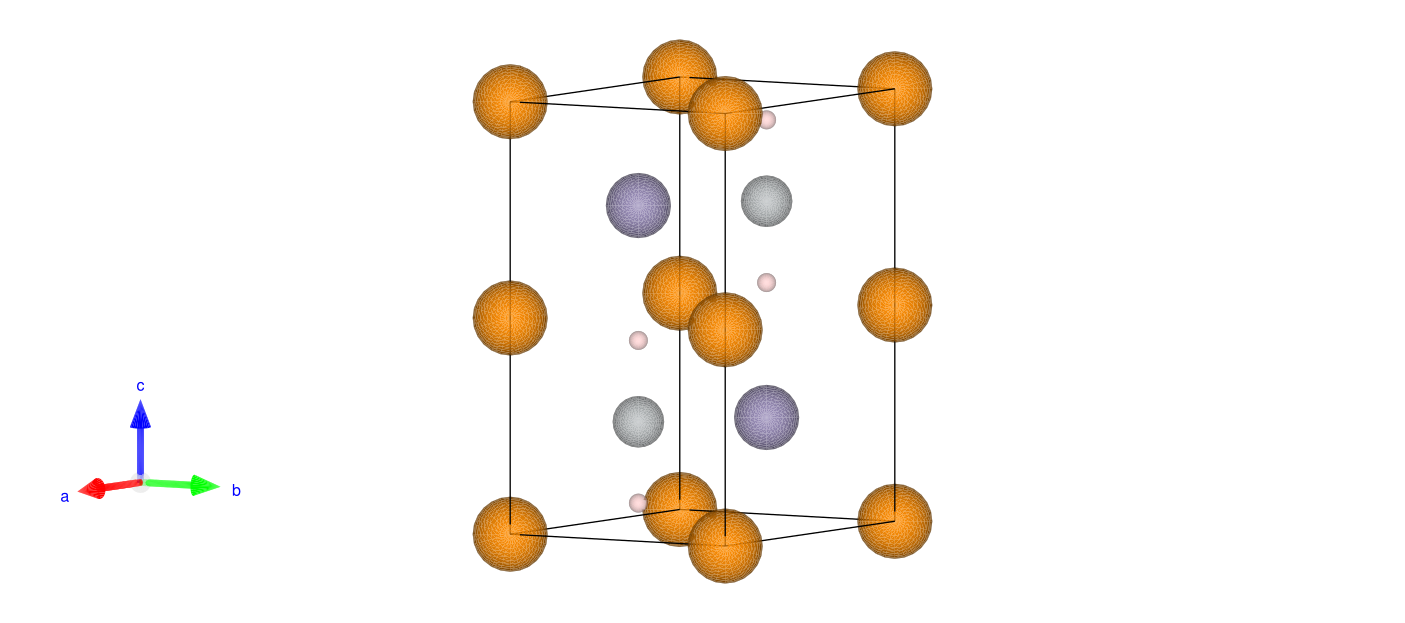}
        \subcaption{NdNiSnH$_2$ (ground truth)}
    \end{minipage}
    \begin{minipage}[c]{0.24\textwidth}
        \centering
        \includegraphics[width=\textwidth]{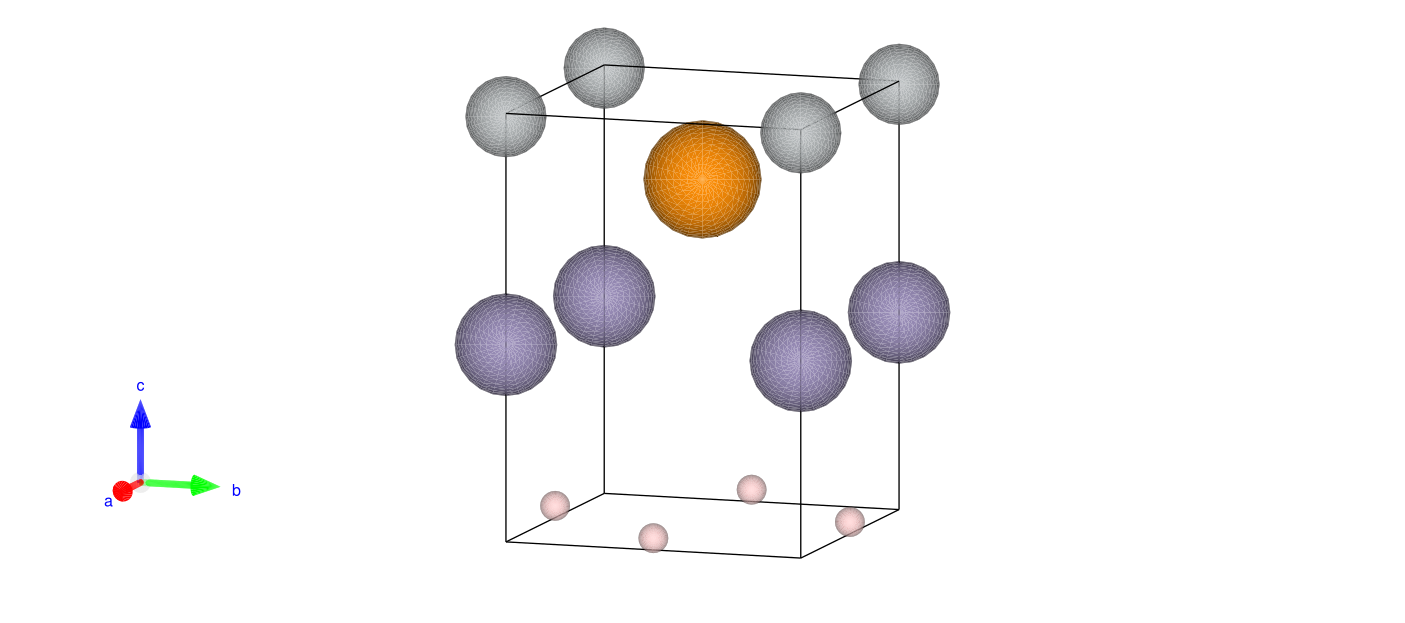}
        \subcaption{NdNiSnH$_2$ (predicted)}
    \end{minipage}\\
     \caption{\textbf{Sample structure prediction by ParetoCSP.}\\Every ground truth structure is followed by the predicted structure. (a) - (p) shows that the structures of MnAl, ZrSe$_2$, GeMo$_3$, SrTiO$_3$, Ta$_2$N$_3$O, and GaBN$_2$ were successfully predicted, while (q) - (t) shows that ParetoCSP was unable to predict the structures of GaSeCl, and NdNiSnH$_2$. All the structures were visualized using VESTA. For better visualization, we set the fractional coordinate ranges of all axis to a maximum of $3$ for Ta$_2$N$_3$O, GaBN$_2$, and GaSeCl, and we used the space-filling style for Ta$_2$N$_3$O, and GaSeCl. Besides these, we set the fractional coordinate ranges of all axis to a maximum of $1$ for all structures, and used the ball-and-stick style.}
     \label{fig:example_comp}
\end{figure}

We observed that ParetoCSP successfully found the most stable structures of all cubic and hexagonal binary crystals and most tetragonal binary crystals in the benchmark dataset. The three unsuccessful binary crystals that ParetoCSP failed to identify their exact structures are Ga$_2$Te$_3$ (monoclinic), Li$_2$Al (orthorhombic), and Ba$_2$Hg (tetragonal). For ternary crystals, ParetoCSP successfully determined the exact stable structures for all tetragonal crystals and most cubic and hexagonal crystals. However, there were four instances where the prediction failed, namely for Li$_2$ZnSi (hexagonal), Cd$_2$AgPt (cubic), GaSeCl (orthorhombic), and K$_2$PdS$_2$ (orthorhombic). In the case of quarternary crystals, ParetoCSP achieved dominance over most hexagonal and tetragonal structures. Li$_2$MgCdP$_2$ (tetragonal), Sr$_2$BBrN$_2$ (trigonal), ZrCuSiAs (tetragonal), NdNiSnH$_2$ (hexagonal), MnCoSnRh (cubic), Mg$_2$ZnB$_2$Ir$_5$ (tetragonal), Ba$_2$CeTaO$_6$ (monoclinic) are the seven quarternary failure cases for ParetoCSP in terms of finding exact structures. Based on these observations, we can claim that ParetoCSP combined with M3GNet IAP demonstrated notable efficacy in predicting cubic, hexagonal, and tetragonal crystalline materials. However, its performance in predicting monoclinic and orthorhombic crystals is comparatively less successful. This can be accounted due to the higher number of degrees of freedom of monoclinic and orthorhombic crystal systems compared to simpler crystal systems like cubic or hexagonal. Also monoclinic and orthorhombic crystals have a varied range of complex structural motifs, which makes CSP algorithms difficult to predict their exact structures. However, this does not diminish the claim that our algorithm is the best among the four ML potential based CSP algorithms evaluated here. Later, we demonstrated that the other CSP algorithms also faced similar challenges. Ground truth and predicted structures of sample crystals are shown in Fig.~\ref{fig:example_comp} using the VESTA tool, which contains examples of both successful and unsuccessful predictions.

Now, we analyze the performance of ParetoCSP in terms of the quantitative performance metrics. As mentioned before, we used a set of seven performance metrics to evaluate the prediction performance of different CSP algorithms. The values of each performance metrics for all $55$ chosen crystal is shown in Table~\ref{table:result_m3gnet}. Ideally, all the performance metric values should be zero if the predicted structure and the ground truth structure are exactly the same. We identified the values of the failure cases which indicate the \textit{poor quality} of the predictions. The process for determining them involved identifying the highest value for each performance metric among all successful predictions (we name them \textit{satisfactory} values), and then selecting the values that exceeded those for the failed predictions. We have highlighted these values in bold letters in Table~\ref{table:result_m3gnet}. We noticed that with the exception of K$_2$PdS$_2$ and ZrCuSiAs, all but $12$ of the failed cases demonstrated higher energy distance values compared to the satisfactory energy distance value ($0.7301$ eV/atom), indicating non-optimal predicted structures. Similarly, for Sinkhorn distance (SD), apart from ZrCuSiAs, the remaining $13$ unsuccessful predictions exhibited significantly higher values than the satisfactory SD value ($5.6727$\AA), suggesting poor prediction quality. For W$_{rmse}$ and W$_{mae}$, we assigned a cross ($\times$) to indicate if the predicted structure and the target structure do not have similar wyckoff position configurations in the symmetrized structures and thus they cannot be calculated. We observed that, $11$ out of $14$ failed predictions (symmetrized) do not have a similar wyckoff position compared to the ground truth symmetrized structure, indicating unsuccessful predictions. However, for Chamfer distance (CD) metric, only $6$ out of $14$ failed predictions displayed higher quantities than the satisfactory CD value ($3.8432$\AA), indicating that CD was not the most suitable metric for measuring prediction quality in crystal structures for our algorithm. In contrast, Hausdorff distance (HD) showed that $10$ out of $14$ failed predictions had higher values than the satisfactory HD value ($3.7665$\AA). Notably, the only performance metric that consistently distinguished between optimal and non-optimal structures across all failed predictions is crystal fingerprint (FP) metric (satisfactory value: $0.9943$), demonstrating its effectiveness in capturing the differences between these structures. In conclusion, all the metrics provided strong evidence of the non-optimal nature of the $14$ failed structures.

\renewcommand{\arraystretch}{1.4}
\begin{longtable}{l c c c c}
\caption{\textbf{Performance comparison of ParetoCSP with baseline algorithms.}\\Successful and failed predictions via manual inspection are denoted by a \cmark and \xmark, respectively. ParetoCSP with M3GNet achieved the highest success rate in finding the exact structures of these crystals, GN-OA with M3GNet achieved the second best success rate. ParetoCSP with MEGNet performed as the third-best, while GN-OA with MEGNet performed the poorest. These results highlight the significant impact of using M3GNet IAP as crystal final energy predictor and structure relaxer, and the effectiveness of the AFPO-based GA as a structure search function.}
\label{table:result_comparison}\\
\hline\hline
\textbf{Composition} & \multicolumn{1}{l}{\begin{tabular}[c]{@{}c@{}}\textbf{ParetoCSP}\\\textbf{with M3GNet (Default)}\end{tabular}} & \multicolumn{1}{l}{\begin{tabular}[c]{@{}c@{}}\textbf{ParetoCSP}\\\textbf{with MEGNet}\end{tabular}} & \multicolumn{1}{l}{\begin{tabular}[c]{@{}c@{}}\textbf{GN-OA}\\\textbf{with M3GNet}\end{tabular}} & \multicolumn{1}{l}{\begin{tabular}[c]{@{}c@{}}\textbf{GN-OA}\\\textbf{with MEGNet (Default)}\end{tabular}} \\ 
\hline\hline
TiCo & \cmark & \cmark & \cmark & \xmark \\
CrPd$_3$ & \cmark & \cmark & \xmark & \xmark \\
GaNi$_3$ & \cmark & \cmark & \cmark & \cmark \\
ZrSe$_2$ & \cmark & \cmark & \cmark & \cmark \\
MnAl & \cmark & \cmark & \cmark & \cmark \\
NiS$_2$ & \cmark & \xmark & \cmark & \cmark \\
TiO$_2$ & \cmark & \cmark & \cmark & \cmark  \\
NiCl & \cmark & \xmark & \xmark & \xmark  \\
AlNi$_3$ & \cmark & \cmark & \cmark & \cmark  \\
CuBr & \cmark & \xmark & \xmark & \xmark  \\
VPt$_3$ & \cmark & \cmark & \cmark & \cmark  \\
MnCo & \cmark & \cmark & \cmark & \cmark  \\
BN & \cmark & \cmark & \cmark & \cmark  \\
GeMo$_3$ & \cmark & \cmark & \cmark & \cmark  \\
Ca$_3$V & \cmark & \cmark & \xmark & \xmark  \\
Ga$_2$Te$_3$ & \xmark & \xmark & \xmark & \xmark \\
CoAs$_2$ & \cmark & \xmark & \xmark & \xmark  \\
Li$_2$Al & \xmark & \xmark & \xmark & \xmark  \\
VS & \cmark & \xmark & \cmark & \xmark  \\
Ba$_2$Hg & \xmark & \xmark & \xmark & \xmark \\
\hline

SrTiO$_3$ & \cmark & \cmark & \cmark & \cmark  \\
Al$_2$FeCo & \cmark & \cmark & \xmark & \xmark \\
GaBN$_2$ & \cmark & \cmark & \xmark & \xmark  \\
AcMnO$_3$ & \cmark & \cmark & \cmark & \cmark  \\
PaTlO$_3$ & \cmark & \cmark & \cmark & \cmark  \\
CdCuN & \cmark & \xmark & \xmark & \xmark  \\
HoHSe & \cmark & \xmark & \cmark & \xmark  \\
Li$_2$ZnSi & \xmark & \xmark & \xmark & \xmark  \\
Cd$_2$AgPt & \xmark & \xmark & \xmark & \xmark  \\
AlCrFe$_2$ & \cmark & \xmark & \xmark & \xmark \\
ZnCdPt$_2$ & \cmark & \xmark & \xmark & \xmark \\
EuAlSi & \cmark & \xmark & \cmark & \cmark \\
Sc$_3$TlC & \cmark & \cmark & \cmark & \cmark \\
GaSeCl & \xmark & \xmark & \xmark & \xmark  \\
CaAgN & \cmark & \xmark & \cmark & \xmark  \\
BaAlGe & \cmark & \cmark & \cmark & \xmark \\
K$_2$PdS$_2$ & \xmark & \xmark & \xmark & \xmark  \\
KCrO$_2$ & \cmark & \xmark & \xmark & \xmark  \\
TiZnCu$_2$ & \cmark & \cmark & \cmark & \cmark  \\
Ta$_2$N$_3$O & \cmark & \xmark & \xmark & \xmark  \\
\hline

AgBiSeS & \cmark & \cmark & \xmark & \xmark  \\
ZrTaNO & \cmark & \xmark & \cmark & \xmark  \\
MnAlCuPd & \cmark & \xmark & \xmark & \xmark  \\
CsNaICl & \cmark & \xmark & \cmark & \xmark  \\
DyThCN & \cmark & \xmark & \cmark & \xmark  \\
Li$_2$MgCdP$_2$ & \xmark & \xmark & \xmark & \xmark  \\
SrWNO$_2$ & \cmark & \xmark & \xmark & \xmark  \\
Sr$_2$BBrN$_2$ & \xmark & \xmark & \xmark & \xmark  \\
ZrCuSiAs & \xmark & \xmark & \xmark & \xmark  \\
NdNiSnH$_2$ & \xmark & \xmark & \xmark & \xmark  \\
MnCoSnRh & \xmark & \xmark & \xmark & \xmark  \\
Mg$_2$ZnB$_2$Ir$_5$ & \xmark & \xmark & \xmark & \xmark  \\
AlCr$_4$GaC$_2$ & \cmark & \cmark & \xmark & \xmark  \\
Y$_3$Al$_3$NiGe$_2$ & \cmark & \xmark & \xmark & \xmark  \\
Ba$_2$CeTaO$_6$ & \xmark & \xmark & \xmark & \xmark  \\
\hline\hline
Accuracy & \multicolumn{1}{l}{\begin{tabular}[c]{@{}c@{}}Overall: \boldmath{$74.55\%$}\\Binary: \boldmath{$85\%$}\\Ternary: \boldmath{$80\%$}\\Quarternary: \boldmath{$53.333\%$}\end{tabular}} & \multicolumn{1}{l}{\begin{tabular}[c]{@{}c@{}}Overall: \boldmath{$40\%$}\\Binary: \boldmath{$60\%$}\\Ternary: \boldmath{$40\%$}\\Quarternary: \boldmath{$13.333\%$}\end{tabular}} & \multicolumn{1}{l}{\begin{tabular}[c]{@{}c@{}}Overall: \boldmath{$43.636\%$}\\Binary: \boldmath{$60\%$}\\Ternary: \boldmath{$45\%$}\\Quarternary: \boldmath{$20\%$}\end{tabular}} & \multicolumn{1}{l}{\begin{tabular}[c]{@{}c@{}}Overall: \boldmath{$29.091\%$}\\Binary: \boldmath{$50\%$}\\Ternary: \boldmath{$30\%$}\\Quarternary: \boldmath{$0\%$}\end{tabular}} \\
\hline\hline
\end{longtable}

\FloatBarrier

\renewcommand{\arraystretch}{1.4}
\begin{longtable}{l c c c c c c c}
\caption{\textbf{Quantitative performance metrics of ParetoCSP with M3GNet for the \boldmath{$55$} benchmark crystals evaluated in this work}.\\For each metric and each failure cases, the values which are greater than the range of exact predictions are denoted by bold letters to mark as high values that quantitatively shows their non-optimality. Binary, ternary, and quarternary crystals are separated by single horizontal lines.}
\label{table:result_m3gnet}\\
\hline\hline
\textbf{Crystal} & \textbf{ED} & \boldmath{$W_{rmse}$} & \boldmath{$W_{mae}$} & \textbf{SD} & \textbf{CD} & \textbf{HD} & \textbf{FP}\\ 
\hline\hline
TiCo & $0.0009$ & $0.0$ & $0.0$ & $0.007$ & $0.007$ & $0.007$ & $0.0$ \\
CrPd$_3$ & $0.0071$ & $0.0$ & $0.0$ & $0.0408$ & $0.0204$ & $0.0136$ & $0.0$ \\ 
GaNi$_3$ & $0.0355$ & $0.0$ & $0.0$ & $0.0839$ & $0.042$ & $0.028$ & $0.0$ \\ 
ZrSe$_2$ & $0.0206$ & $0.0062$ & $0.0025$ & $0.6353$ & $0.4235$ & $0.5848$ & $0.3243$ \\ 
MnAl & $0.0$ & $0.0$ & $0.0$ & $0.0002$ & $0.0002$ & $0.0002$ & $0.0$ \\ 
NiS$_2$ & $0.2016$ & $0.2889$ & $0.2303$ & $5.6727$ & $3.8432$ & $3.7665$ & $0.269$ \\ 
TiO$_2$ & $0.6931$ & $0.2304$ & $0.1431$ & $4.209$ & $2.8535$ & $1.8551$ & $0.9793$\\ 
NiCl & $0.3284$ & $0.2562$ & $0.1723$ & $1.3811$ & $2.3407$ & $1.1495$ & $0.6431$ \\ 
AlNi$_3$ & $0.0234$ & $0.0$ & $0.0$ & $0.0727$ & $0.0363$ & $0.0242$ & $0.0$ \\ 
CuBr & $0.3225$ & $0.2521$ & $0.1784$ & $1.8724$ & $2.5043$ & $1.0065$ & $0.3054$ \\ 
VPt$_3$ & $0.2415$ & $0.3235$ & $0.2411$ & $1.3424$ & $0.2395$ & $0.2805$ & $0.1772$ \\ 
MnCo & $0.0$ & $0.0$ & $0.0$ & $0.0001$ & $0.0001$ & $0.0001$ & $0.0$ \\ 
BN & $0.3643$ & $0.4026$ & $0.2454$ & $2.513$ & $1.947$ & $2.608$ & $0.8948$ \\ 
GeMo$_3$ & $0.0401$ & $0.0$ & $0.0$ & $0.1894$ & $0.0473$ & $0.0325$ & $0.0$ \\  
Ca$_3$V & $0.4592$ & $0.2048$ & $0.1149$ & $3.3111$ & $2.8356$ & $3.6542$ & $0.019$ \\
Ga$_2$Te$_3$ & \boldmath{$2.0112$} & \boldmath{$\times$} & \boldmath{$\times$} & \boldmath{$53.3896$} & \boldmath{$4.6825$} & \boldmath{$4.8998$} & \boldmath{$1.7875$} \\ 
CoAs$_2$ & $0.4629$ & $0.4389$ & $0.2684$ & $5.3617$ & $2.8407$ & $2.9208$ & $0.9943$ \\ 
Li$_2$Al & \boldmath{$30.7051$} & \boldmath{$\times$} & \boldmath{$\times$} & \boldmath{$61.9154$} & \boldmath{$3.9575$} & \boldmath{$4.8314$} & \boldmath{$2.1345$} \\ 
VS & $0.4204$ & $0.2477$ & $0.1806$ & $1.9372$ & $1.3665$ & $1.8303$ & $0.9189$ \\ 
Ba$_2$Hg & \boldmath{$5.206$} & \boldmath{$\times$} & \boldmath{$\times$} & \boldmath{$8.7511$} & \boldmath{$4.9936$} & \boldmath{$7.3342$} & \boldmath{$1.2468$} \\ 
\hline

SrTiO$_3$ & $0.0185$ & $0.0$ & $0.0$ & $0.0934$ & $0.0374$ & $0.0271$ & $0.0$ \\ 
Al$_2$FeCo & $0.0098$ & $0.2357$ & $0.112$ & $0.137$ & $0.0685$ & $0.0658$ & $0.1755$ \\ 
GaBN$_2$ & $0.0041$ & $0.3889$ & $0.289$ & $2.1663$ & $1.5589$ & $1.9171$ & $0.0455$ \\ 
AcMnO$_3$ & $0.0385$ & $0.0$ & $0.0$ & $0.116$ & $0.0464$ & $0.0336$ & $0.0$ \\ 
PaTlO$_3$ & $0.0136$ & $0.0$ & $0.0$ & $0.0924$ & $0.037$ & $0.0268$ & $0.0$ \\ 
CdCuN & $0.0031$ & $0.441$ & $0.4259$ & $2.7337$ & $2.9172$ & $2.2949$ & $0.0397$ \\
HoHSe & $0.0033$ & $0.3643$ & $0.3148$ & $2.859$ & $1.906$ & $1.9716$ & $0.0575$ \\ 
Li$_2$ZnSi & \boldmath{$25.3593$} & \boldmath{$\times$} & \boldmath{$\times$} & \boldmath{$34.3079$} & $2.9587$ & \boldmath{$4.104$} & \boldmath{$1.8731$} \\ 
Cd$_2$AgPt & \boldmath{$22.5447$} & \boldmath{$\times$} & \boldmath{$\times$} & \boldmath{$16.9997$} & $3.5895$ & \boldmath{$4.2417$} & \boldmath{$2.4137$} \\ 
AlCrFe$_2$ & $0.6621$ & $0.2486$ & $0.1507$ & $3.6931$ & $2.2245$ & $2.2518$ & $0.7886$ \\
ZnCdPt$_2$ & $0.0384$ & $0.4717$ & $0.4503$ & $3.2733$ & $3.5537$ & $2.0384$ & $0.0643$ \\
EuAlSi & $0.0495$ & $0.3849$ & $0.2963$ & $4.5051$ & $3.0034$ & $2.2451$ & $0.3419$ \\
Sc$_3$TlC & $0.0026$ & $0.0$ & $0.0$ & $0.0431$ & $0.0173$ & $0.0125$ & $0.0$ \\ 
GaSeCl & \boldmath{$23.3337$} & \boldmath{$\times$} & \boldmath{$\times$} & \boldmath{$38.0257$} & \boldmath{$8.615$} & \boldmath{$11.7449$} & \boldmath{$2.0172$} \\ 
CaAgN & $0.0064$ & $0.441$ & $0.4259$ & $3.6479$ & $3.1055$ & $2.4023$ & $0.0483$ \\ 
BaAlGe & $0.002$ & $0.4547$ & $0.3889$ & $3.0476$ & $1.6942$ & $2.5291$ & $0.0326$ \\  
K$_2$PdS$_2$ & $0.5466$ & $0.2467$ & $0.1377$ & \boldmath{$22.0109$} & $3.7687$ & $3.5226$ & \boldmath{$1.3316$} \\ 
KCrO$_2$ & $0.0342$ & $0.2740$ & $0.1934$ & $2.5233$ & $1.9562$ & $1.8946$ & $0.6105$ \\ 
TiZnCu$_2$ & $0.0188$ & $0.4083$ & $0.3344$ & $3.8363$ & $2.83$ & $1.609$ & $0.6861$ \\ 
Ta$_2$N$_3$O & $0.4603$ & $0.2357$ & $0.1111$ & $3.144$ & $2.3813$ & $1.4458$ & $0.7499$ \\ 
\hline

AgBiSeS & $0.0154$ & $0.0$ & $0.0$ & $0.1914$ & $0.0957$ & $0.0808$ & $0.1298$ \\
ZrTaNO & $0.0935$ & $0.5182$ & $0.5$ & $0.4704$ & $0.2352$ & $0.2191$ & $0.4131$ \\ 
MnAlCuPd & $0.0187$ & $0.1719$ & $0.0865$ & $3.3567$ & $2.3023$ & $2.219$ & $0.7371$ \\
CsNaICl & $0.0046$ & $0.5$ & $0.5$ & $0.1822$ & $0.0911$ & $0.0848$ & $0.1639$ \\
DyThCN & $0.0322$ & $0.4082$ & $0.3333$ & $0.1057$ & $0.0529$ & $0.0451$ & $0.0216$ \\
Li$_2$MgCdP$_2$ & \boldmath{$39.8356$} & \boldmath{$\times$} & \boldmath{$\times$} & \boldmath{$36.702$} & $3.4202$ & \boldmath{$4.3517$} & \boldmath{$1.8915$} \\ 
SrWNO$_2$ & $0.0378$ & $0.0$ & $0.0$ & $0.2707$ & $0.1083$ & $0.1001$ & $0.0867$ \\
Sr$_2$BBrN$_2$ & \boldmath{$10.728$} & \boldmath{$\times$} & \boldmath{$\times$} & \boldmath{$34.7446$} & $2.9484$ & \boldmath{$4.7848$} & \boldmath{$1.0966$} \\ 
ZrCuSiAs & $0.1566$ & $0.2459$ & $0.1411$ & $5.63$ & $1.4075$ & $1.5158$ & \boldmath{$1.7131$} \\
NdNiSnH$_2$ & \boldmath{$24.8101$} & $0.4252$ & $0.2993$ & \boldmath{$10.3403$} & $3.4393$ & $3.6793$ & \boldmath{$1.945$} \\ 
MnCoSnRh & \boldmath{$56.8397$} & \boldmath{$\times$} & \boldmath{$\times$} & \boldmath{$12.3179$} & $3.0676$ & $3.5955$ & \boldmath{$1.2971$} \\ 
Mg$_2$ZnB$_2$Ir$_5$ & \boldmath{$6.8128$} & \boldmath{$\times$} & \boldmath{$\times$} & \boldmath{$60.6003$} & \boldmath{$6.7022$} & \boldmath{$7.5961$} & \boldmath{$1.5616$} \\ 
AlCr4GaC2 & $0.0234$ & $0.5563$ & $0.3984$ & $4.9214$ & $2.4287$ & $1.7347$ & $0.0986$ \\ 
Y$_3$Al$_3$NiGe$_2$ & $0.7301$ & $0.1035$ & $0.057$ & $4.0638$ & $3.1641$ & $2.9705$ & $0.5302$ \\
Ba$_2$CeTaO$_6$ & \boldmath{$52.5924$} & \boldmath{$\times$} & \boldmath{$\times$} & \boldmath{$78.9662$} & \boldmath{$5.3529$} & \boldmath{$6.8904$} & \boldmath{$1.9963$} \\ 
\hline\hline
\end{longtable}

\subsection{Performance comparison with GN-OA}\label{subsec: comp2}
As reported in ~\cite{gnoa}, the GN-OA algorithm achieved the highest performance when utilizing Bayesian Optimization (BO)~\cite{tpe} as the optimization algorithm and MEGNet neural network model as the formation energy predictor to guide the optimization process (default GN-OA). Based on the data presented in Table~\ref{table:result_comparison}, we observed that GN-OA showed a significantly lower success rate than that of ParetoCSP. In comparison to ParetoCSP, GN-OA achieved an accuracy of only $50\%$ ($10$ out of $20$ crystals) in predicting structures of binary crystals, whereas ParetoCSP achieved $85\%$ accuracy. For ternary crystals, GN-OA achieved a success rate of $30\%$ ($6$ out of $20$ crystals) compared to ParetoCSP's $80\%$. In the case of quarternary crystals, GN-OA did not achieve a single success, whereas ParetoCSP achieved a success rate of $53.333\%$. Overall, the success rate of GN-OA was only $29.091\%$, which is approximately $2.562$ times lower than the accuracy achieved by ParetoCSP. Moreover, GN-OA could not predict any structure that ParetoCSP could not predict. These clearly establish the dominance of ParetoCSP over GN-OA, highlighting the higher quality of structure searching provided by AFPO-based GA compared to BO, and the effectiveness of M3GNet IAP-based final energy prediction compared to MEGNet's formation energy prediction.

To understand the deteriorated performance of GN-OA in our benchmark study, firstly, we found that the CSP experiments conducted in the original study of GN-OA\cite{gnoa} primarily focused on small binary crystals, particularly those with a $1$:$1$ atoms ratio. Secondly, a majority of these binary crystals belonged to four groups, namely oxide, sulfide, chloride, and fluoride, that demonstrates the lack of diversity in the GN-OA's benchmark set (see Supplementary Fig. S1b). Moreover, most of the crystals examined had the cubic crystal system (mostly belonging to the $Fm-3m$ space group). It merely explored other crystal systems or space group. This choice of test structures for experimentation was insufficient in terms of CSP where only a few crystals possess all these specific properties. A more thorough exploration of diverse crystal systems and space groups was necessary to demonstrate GN-OA's CSP performance. Our study effectively demonstrated that the optimization algorithms used in GN-OA are inadequate for predicting more complex crystals (such as quarternary crystals). Furthermore, our empirical findings highlighted the shortcomings of using MEGNet as formation energy predictor in guiding the optimization algorithm towards the optimal crystal structures. In summary, we established that ParetoCSP outperformed GN-OA by achieving a staggering $256.2\%$ higher performance in terms of success rates than that of GN-OA, and the AFPO-based multi-objective GA proved to be a much better structure search algorithm than BO. Additionally, M3GNet IAP provided more accurate energy estimations for effective CSP compared to the MEGNet used in GN-OA. ParetoCSP also performs a further structure refinement using M3GNet IAP after obtaining the final optimized structure from the GA, which contributed to its higher accuracy compared to GN-OA where this is entirely absent.

Fig.~\ref{fig:success} shows performance metric value comparison for some sample crystals. For better visualization, we limited the $y$-axis values to $20$ for Fig.~\ref{fig: success_ed} and \ref{fig: success_sd}, and to $10$ for Fig.~\ref{fig: success_cd} and \ref{fig: success_hd}. We found that the default ParetoCSP with M3GNet achieved lower (better) performance metric values for all the chosen sample crystals in terms of the metrics of ED, HD, and FP and for the majority of the cases for SD, and CD, compared to the default GN-OA. For some crystals (e.g., Ta$_2$N$_3$O, AgBiSeS, MnAlCuPd, SrWNO$_2$) the differences in the performance metric quantities are huge, indicating ParetoCSP's strong dominance over the default GN-OA.

\subsection{Performance comparison of CSP algorithms with different energy models}\label{subsec: comp3}
As discussed in the previous section, M3GNet universal IAP proved to be a better energy predictor than MEGNet. To fairly and objectively evaluate and compare our algorithm's performance, we replaced ParetoCSP's final energy calculator (M3GNet) with the MEGNet GNN for formation energy evaluation. Subsequently, we also replace MEGNet with M3GNet in GN-OA to show that the M3GNet IAP performs better than MEGNet for predicting the most stable energy for CSP. As a result, we ran experiments on four algorithms - ParetoCSP with M3GNet (default ParetoCSP), ParetoCSP with MEGNet, GN-OA with MEGNet (default GN-OA), and GN-OA with M3GNet.

\begin{figure}[ht] 
    \centering
    \begin{minipage}[c]{0.495\textwidth}
        \centering
        \includegraphics[width=\textwidth]{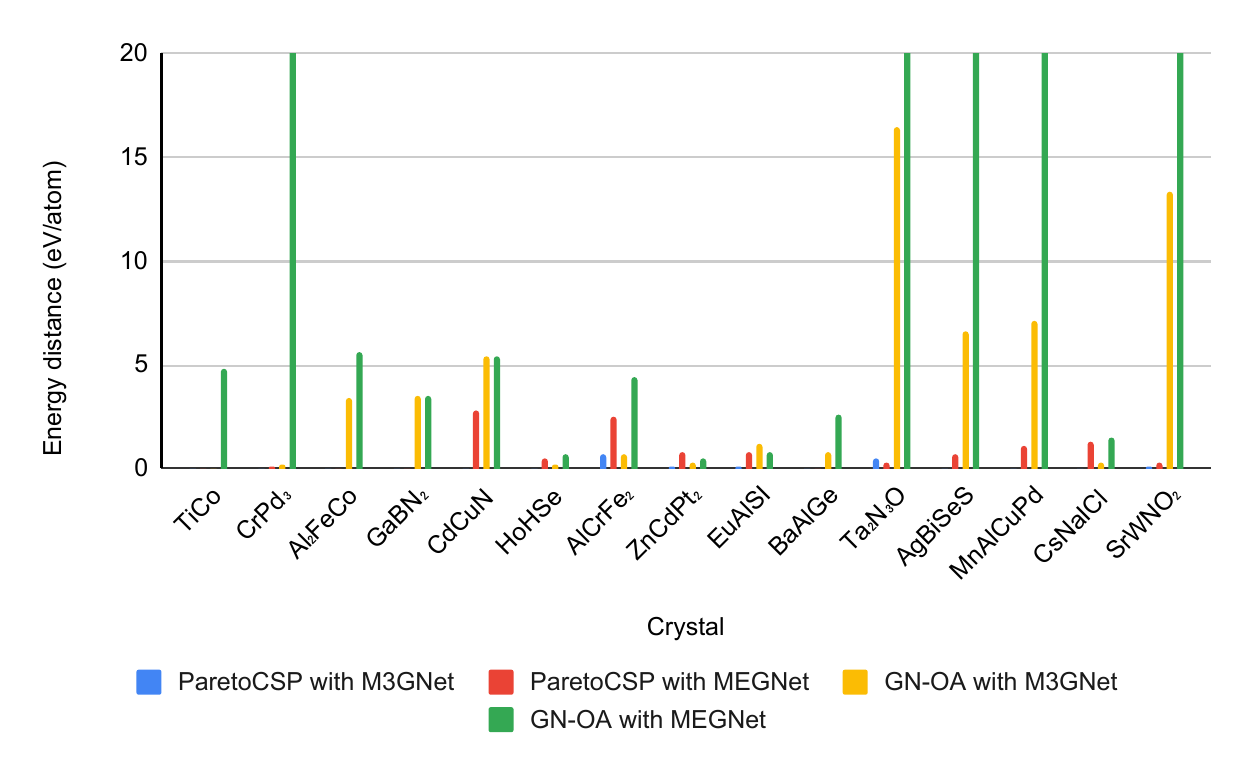}
        \subcaption{Energy distance}
        \label{fig: success_ed}
    \end{minipage}
    \begin{minipage}[c]{0.495\textwidth}
        \centering
        \includegraphics[width=\textwidth]{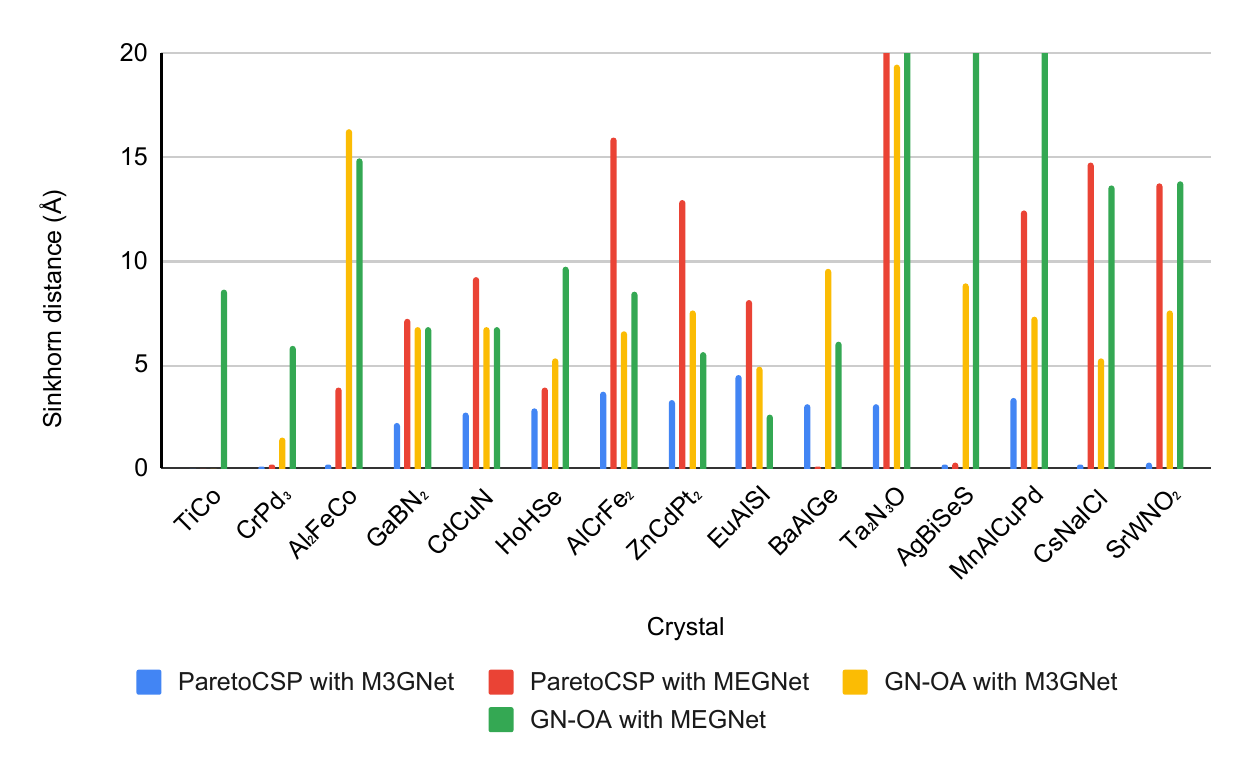}
        \subcaption{Sinkhorn distance}
        \label{fig: success_sd}
    \end{minipage}\\

    \begin{minipage}[c]{0.495\textwidth}
        \centering
        \includegraphics[width=\textwidth]{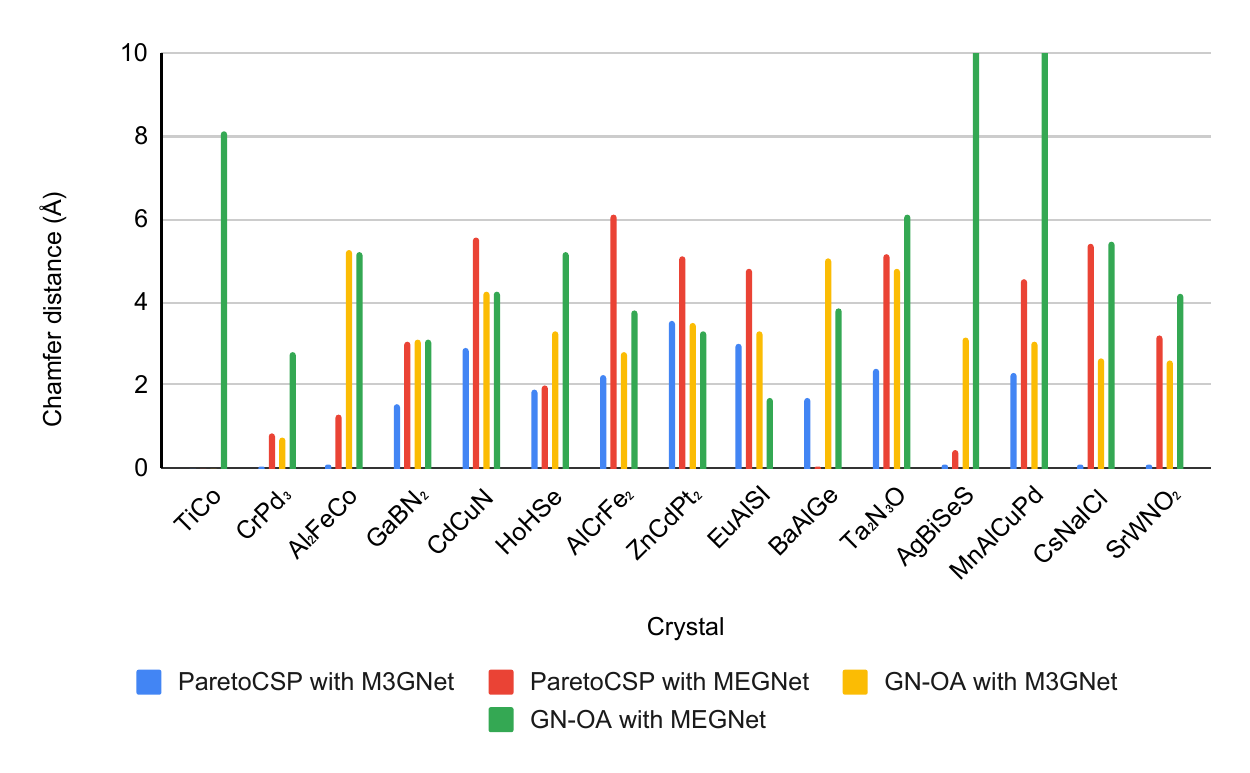}
        \subcaption{Chamfer distance}
        \label{fig: success_cd}
    \end{minipage}
    \begin{minipage}[c]{0.495\textwidth}
        \centering
        \includegraphics[width=\textwidth]{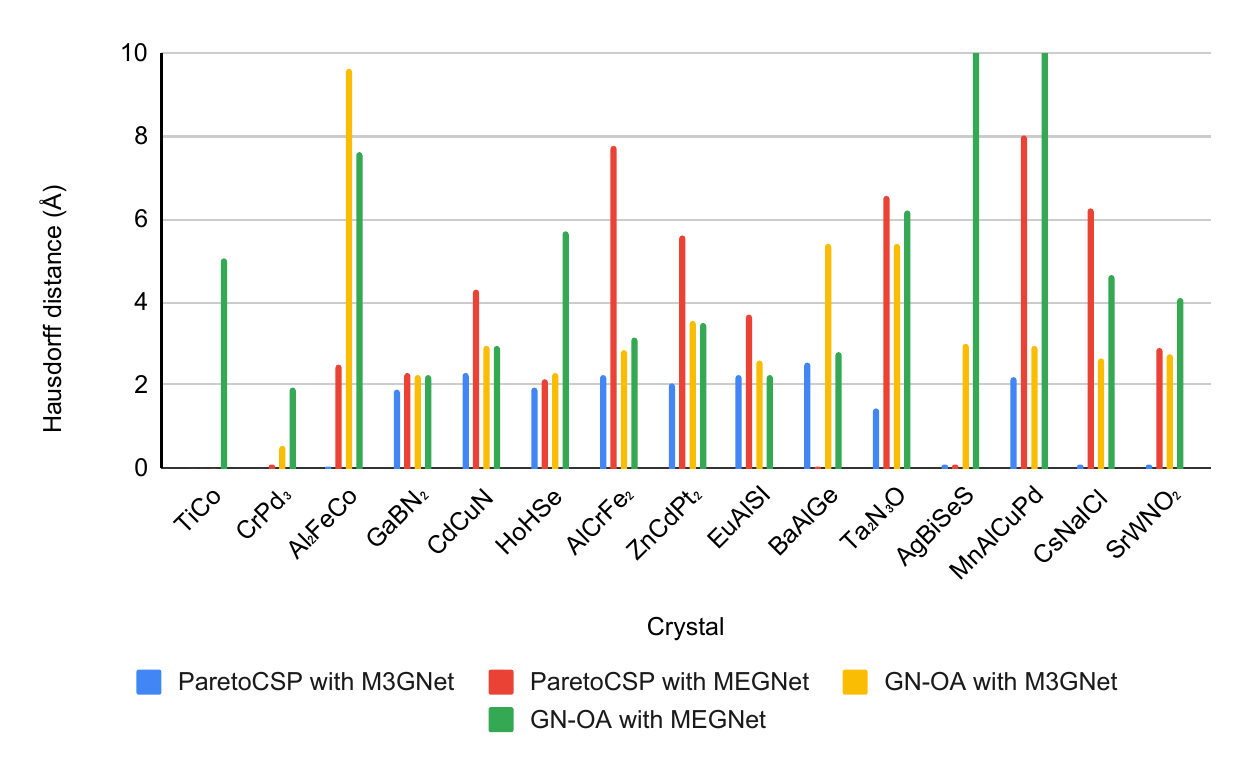}
        \subcaption{Hausdorff distance}
        \label{fig: success_hd}
    \end{minipage}\\

    \begin{minipage}[c]{0.55\textwidth}
        \centering
        \includegraphics[width=\textwidth]{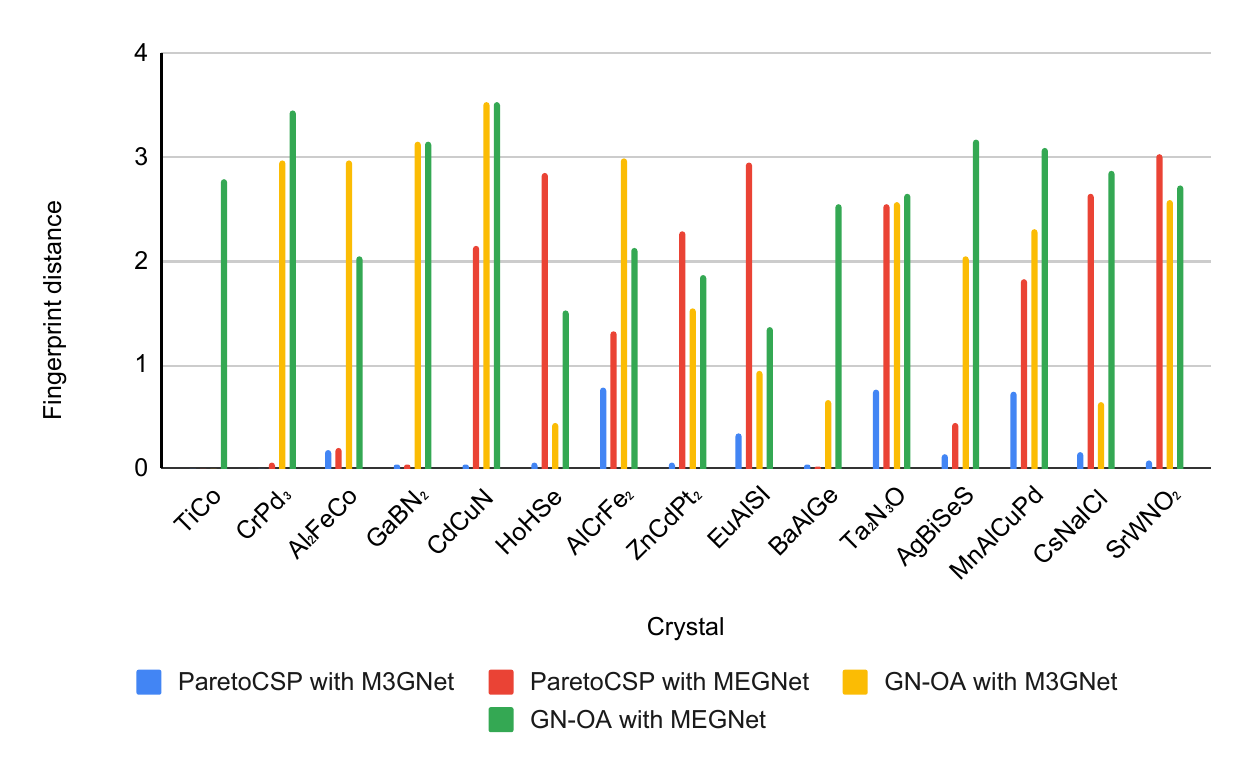}
        \subcaption{Fingerprint distance}
        \label{fig: success_fp}
    \end{minipage}\\
     \caption{\textbf{Performance metric comparison of different CSP algorithms evaluated over the sample benchmark crystals.}\\The metric values of ParetoCSP with M3GNet is much smaller (better) than those of other baseline algorithms, which quantitatively shows its superiority. In most cases, GN-OA with MEGNet's metric values are the highest (worst) which is aligned with the observation that it demonstrated the poorest performance among all CSP algorithms.}
     \label{fig:success}
\end{figure}

The results of ParetoCSP with M3GNet have been discussed in detail in Section~\ref{susec: paretocsp_results}. ParetoCSP with MEGNet outperformed the default GN-OA by a factor of $\approx 1.31$ in terms of exact structure prediction accuracy. Individually, ParetoCSP with MEGNet achieved $60\%$ ($12$ out of $20$), $40\%$ ($8$ out of $20$), and $13.333\%$ ($2$ out of $15$) accuracy in predicting structures of binary, ternary, and quarternary crystals, respectively. In comparison, GN-OA with MEGNet achieved accuracies of $50\%$, $30\%$, and $0\%$ for binary, ternary, and quarternary crystals, respectively. This comparison clearly demonstrated that the AFPO-based GA is a more effective structure search method than BO. NiS$_2$ and EuAlSi are the only two crystals (both hexagonal) that GN-OA with MEGNet could predict the exact structures of but ParetoCSP with MEGNet could not. But the opposite is true for $8$ crystals including GaNi$_3$, GaBN$_2$, BaAlGe, AgBiSeS, etc., predominantly belonging to the tetragonal crystal system. Additionally, ParetoCSP with MEGNet were not successful in predicting any structure that ParetoCSP with M3GNet could not, strongly indicating the necessity for M3GNet as the energy predicting function (outperformed ParetoCSP with MEGNet by a factor of $\approx 1.86$). From Fig.~\ref{fig:success}, we can see that ParetoCSP with M3GNet achieved much lower performance metric values than ParetoCSP with MEGNet for the majority of the cases, indicating its better prediction caliber.

Based on the analysis conducted so far, two hypotheses were formulated: firstly, that GN-OA with M3GNet would outperform the default GN-OA, and secondly, that ParetoCSP with M3GNet would outperform GN-OA with M3GNet. As anticipated, GN-OA with M3GNet outperformed the default GN-OA (by a factor of $\approx 1.5$), again demonstrating M3GNet IAP as a much better energy model than MEGNet. For binary, ternary, and quarternary crystals, respectively, GN-OA with M3GNet (GN-OA with MEGNet) achieved $60\%$ ($50\%$), $35\%$ ($30\%$), and $13.333\%$ ($0\%$), respectively. Moreover, the default GN-OA did not achieve superiority over GN-OA with MEGNet on any chosen crystal, but the opposite is true for $8$ crystals including TiCo, VS, HoHSe, CsNaICl, etc., and a majority of them belongs to the hexagonal crystal system. However, despite the improved performance of GN-OA with M3GNet, it's efficiency still fell short in comparison to ParetoCSP with M3GNet due to the more effective structure search function of the latter, proving both hypothesis true. ParetoCSP with M3GNet outperformed GN-OA with M3GNet by a factor of $\approx 1.71$. Furthermore, the default ParetoCSP accurately predicted every structure that GN-OA with M3GNet successfully predicted. Again from Fig.~\ref{fig:success}, we can see that ParetoCSP with M3GNet achieved smaller performance metric values than GN-OA with M3GNet for the majority of the crystals. In fact, for some crystals such as Al$_2$FeCo, Ta$_2$N$_3$O, AgBiSeS, and SrWNO$_2$, the differences of metric values are enormous. To report the final outcomes, ParetoCSP with M3GNet outperformed all algorithms ($\approx 1.71\times$ the second best, and $\approx 1.86\times$ the third best). GN-OA with M3GNet ranked second best, exceeding the performance of the third best ParetoCSP with MEGNet by a small margin (by a factor of $\approx 1.09$). The default GN-OA demonstrated the lowest performance compared to all other algorithms.

\FloatBarrier

\subsection{Parametric study of ParetoCSP}

As a multi-objective GA, there are several hyper-parameters to set before runing our ParetoCSP algorithm for CSP. Here we conducted experiments with our ParetoCSP algorithm with different parameter settings to evaluate their effect. We selected $8$ crystals for this study containing both successful and unsuccessful predictions, namely TiCo, Ba$_2$Hg, HoHSe, Cd$_2$AgPt, SrTiO$_3$, GaBN$_2$, MnAlCuPd, and AgBiSeS. The hyper-parameters chosen for the study include population size, crossover probability, mutation probability, and total number of generations used. The default parameter set is mentioned in Supplementary Note S1. All the performance results are presented in Table~\ref{table:result_parameter}.

First, we examined the effect of different population sizes on the selected crystals. We ran the experiments with five different population sizes. The results in Table~\ref{table:result_parameter} shows that our algorithm performed best with a population size of $100$. Conversely, it could not accurately predict the structures of any crystal with a population size of $30$, except for SrTiO$_3$. ParetoCSP consistently performed poorly for Ba$_2$Hg and Cd$_2$AgPt with every population size, while the results of SrTiO$_3$ showed the opposite trend.

Second, we analyzed the performance of our algorithm with varying crossover probabilities. The results indicated that the best performance was achieved with a probability of $0.8$, and this was the only probability for which ParetoCSP identified the exact structure of MnAlCuPd. Except GaBN$_2$ and AgBiSeS, for all five other crystals, ParetoCSP showed consistent performance with other crossover probabilities. We observed that our algorithm performed well with higher crossover probabilities for GaBN$_2$, and poorly for AgBiSeS with probability $< 0.2$

Next, we evaluated ParetoCSP's performance with different mutation probabilities. and observed that ParetoCSP performed best with a mutation probability of $0.01$. Only MnAlCuPd and AgBiSeS had their exact structure successfully predicted with this mutation probability, while for other crystals except GaBN$_2$, ParetoCSP performed consistently with other probabilities. Our algorithm successfully predicted the structure of GaBN$_2$ for mutation probabilities $\geq 0.01$.

Finally, we ran experiments with different generations to investigate the impact on algorithm performance. In ~\cite{gnoa}, all experiments were run for $5000$ steps for the BO. However, our results from Table~\ref{table:result_parameter} showed that $1000$ generations were sufficient for ParetoCSP to achieve the optimal results for all $8$ crystals. Except for GaBN$_2$, and AgBiSeS, for all five other crystals, ParetoCSP achieved optimal solutions within $250$ generations. We would like to mention that we did not evaluate for $< 250$ generations, so it is possible that ParetoCSP could perform optimally for these crystals even with a smaller number of generations. None of the above mentioned hyper-parameters could accurately predict the ground truth structures of Ba$_2$Hg, and Cd$_2$AgPt.

\begin{table}[!htb]
\begin{center}
\caption{\textbf{Performance results with different hyper-parameters of ParetoCSP with M3GNet.}\\ Pop, CP, MP, and Gen denote population size, crossover probability, mutation probability, and total number of generations, respectively. The best results are achieved for a population size of $100$, a crossover probability of $0.8$, a mutation probability of $0.01$, and a generation number $\geq 1000$. ParetoCSP failed to identify exact structures of Ba$_2$Hg, and Cd$_2$AgPt for all parameter settings tested in this experiment.}
\label{table:result_parameter}
\renewcommand{\arraystretch}{1.4}
\begin{tabular}{||l || c c c c c c c c||}
\cline{2-9}
\multicolumn{1}{c|}{} & \textbf{TiCo} & \textbf{Ba$_2$Hg} & \textbf{HoHSe} & \textbf{Cd$_2$AgPt} & \textbf{SrTiO$_3$} & \textbf{GaBN$_2$} & \textbf{MnAlCuPd} & \textbf{AgBiSeS} \\ 
\hline\hline
Pop $30$ & \xmark & \xmark & \xmark & \xmark & \cmark & \xmark & \xmark & \xmark \\
Pop $60$ & \cmark & \xmark & \cmark & \xmark & \cmark & \xmark & \xmark & \cmark \\
Pop $100$ & \cmark & \xmark & \cmark & \xmark & \cmark & \cmark & \cmark & \cmark \\
Pop $200$ & \cmark & \xmark & \cmark & \xmark & \cmark & \xmark & \xmark & \cmark \\
Pop $300$ & \cmark & \xmark & \cmark & \xmark & \cmark & \xmark & \cmark & \xmark \\
\hline
CP $0.1$ & \cmark & \xmark & \cmark & \xmark & \cmark & \xmark & \xmark & \xmark \\
CP $0.2$ & \cmark & \xmark & \cmark & \xmark & \cmark & \xmark & \xmark & \cmark \\
CP $0.4$ & \cmark & \xmark & \cmark & \xmark & \cmark & \xmark & \xmark & \cmark \\
CP $0.6$ & \cmark & \xmark & \cmark & \xmark & \cmark & \cmark & \xmark & \cmark \\
CP $0.8$ & \cmark & \xmark & \cmark & \xmark & \cmark & \cmark & \cmark & \cmark \\
\hline
MP $0.0001$ & \cmark & \xmark & \cmark & \xmark & \cmark & \xmark & \xmark & \xmark \\
MP $0.001$ & \cmark & \xmark & \cmark & \xmark & \cmark & \xmark & \xmark & \xmark \\
MP $0.01$ & \cmark & \xmark & \cmark & \xmark & \cmark & \cmark & \cmark & \cmark \\
MP $0.1$ & \cmark & \xmark & \cmark & \xmark & \cmark & \cmark & \xmark & \xmark \\
MP $0.5$ & \cmark & \xmark & \cmark & \xmark & \cmark & \cmark & \xmark & \xmark \\
\hline
Gen $250$ & \cmark & \xmark & \cmark & \xmark & \cmark & \xmark & \cmark & \xmark \\
Gen $500$ & \cmark & \xmark & \cmark & \xmark & \cmark & \cmark & \cmark & \xmark \\
Gen $1000$ & \cmark & \xmark & \cmark & \xmark & \cmark & \cmark & \cmark & \cmark \\
Gen $2000$ & \cmark & \xmark & \cmark & \xmark & \cmark & \cmark & \cmark & \cmark \\
Gen $5000$ & \cmark & \xmark & \cmark & \xmark & \cmark & \cmark & \cmark & \cmark \\
\hline\hline
\end{tabular}
\end{center}
\end{table}

\subsection{Failure case study}
ParetoCSP successfully predicted the structures for $41$ out of $55$ benchmark crystals in this research. Here we conducted a further thorough investigation of the $14$ unsuccessful predictions. For this, we calculated performance metric values of these $14$ structures for all four algorithms discussed in this paper and then experimentally showed the quality of each algorithms' output. We excluded the W$_{rmse}$ and W$_{mae}$ for this study as all four algorithms failed to predict these structures accurately. The results are presented in Fig.~\ref{fig:failure} (only two of them are shown here in the main text, and the rest are shown in the Supplementary File.)

The comparison results for energy distance metric (ED) is presented in Supplementary Fig. S2a. We limited the $y$-axis value to $80$ for better visualization. ParetoCSP with M3GNet dominated all other algorithms for ED, achieving the lowest errors for $9$ out of $14$ crystals. ED is related to the final energy difference between the ground truth and the predicted structure, indicating that predicted structures by ParetoCSP are more energetically closer to the target structures' energy than those by other algorithms. The only failure case where the ParetoCSP had the highest ED value among all algorithms was Li$_2$Al. The three performance metrics SD, CD, and HD, are all related to the atomic sites of the ground truth and predicted crystal. ParetoCSP with M3GNet again outperformed all other algorithms, achieving lowest distance scores for a majority of the failure cases, suggesting that the structures predicted by the ParetoCSP algorithms have the closest atomic site configurations compared to the target structures among all algorithms. We presented the results in Supplementary Fig. S2b, Supplementary Fig. S2c, and Fig.~\ref{fig: failure_hd}, respectively with the $y$-axis of Supplementary Fig. S2b limited to $200$ for visualization purposes. Finally for the fingerprint metric (FP), which is related to the crystal atomic site fingerprint, ParetoCSP with M3GNet achieved the lowest distance errors for $11$ out of $14$ crystals among all algorithms, proving better atomic site prediction quality. The results are shown in Fig.~\ref{fig: failure_fp}. Li$_2$Al again is the only crystal where the default ParetoCSP's FP value is the highest among all.

The observation that Li$_2$Al had the highest ED and FP values for ParetoCSP suggests that the combination of AFPO-based GA and M3GNet might not be the optimal choice for predicting this crystal. On the contrary, ParetoCSP with M3GNet achieved $4$ out of $5$, or $5$ out of $5$ lowest performance metric values for Ga$_2$Te$_3$, K$_2$PdS$_2$, Sr$_2$BBrN$_2$, ZrCuSiAs, MnCoSnRh, and Ba$_2$CeTaO$_6$ indicating that we are on the right track to predict structures of these crystals. In summary, each of the performance metrics is related to some specific features of the ground truth crystals, and ParetoCSP with M3GNet outperforms all other algorithms, which indicates that it predicts structures with better quality (more closer to the ground truth structures) than other algorithms despite none of them are exact solutions.

\begin{figure}[ht] 
    \centering
    \begin{minipage}[c]{0.495\textwidth}
        \centering
        \includegraphics[width=\textwidth]{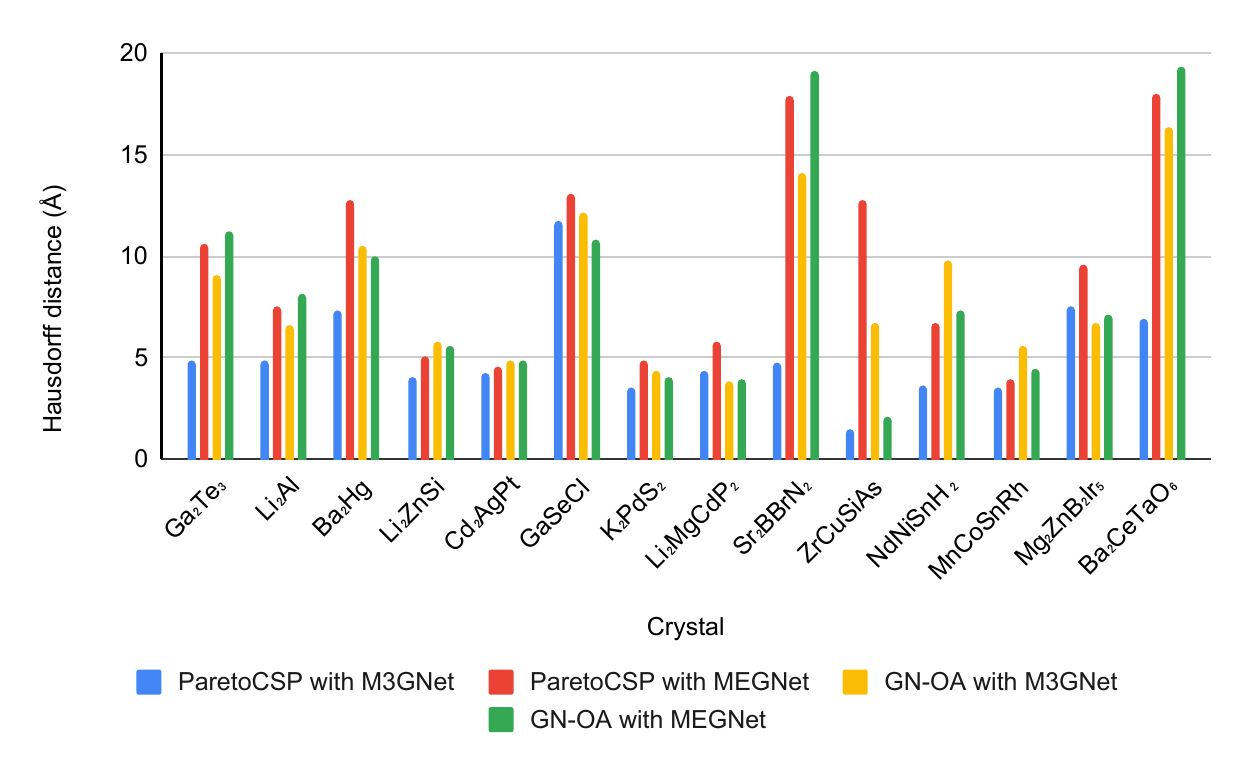}
        \subcaption{Hausdorff distance}
        \label{fig: failure_hd}
    \end{minipage}
    \begin{minipage}[c]{0.495\textwidth}
        \centering
        \includegraphics[width=\textwidth]{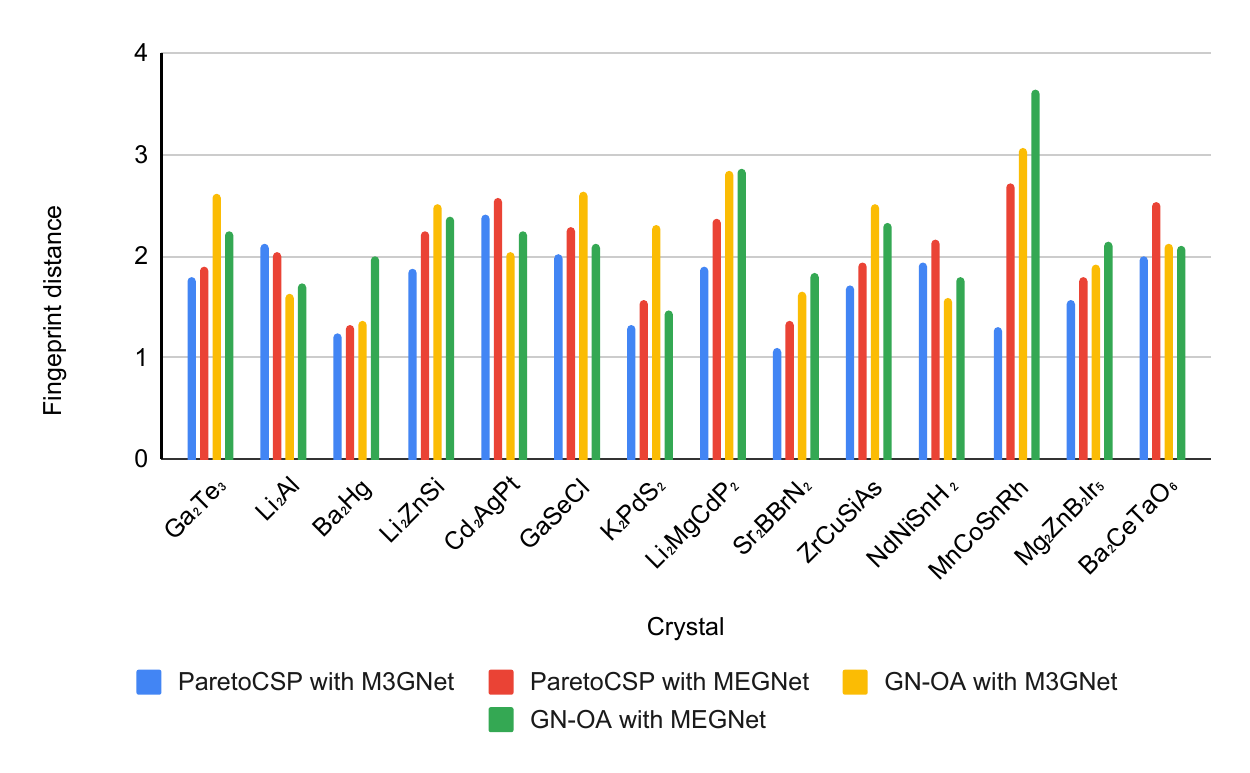}
        \subcaption{Fingerprint distance}
        \label{fig: failure_fp}
    \end{minipage}\\

     \caption{\textbf{Performance metric comparison of structure prediction of different algorithms for the \boldmath{$14$} failure cases of ParetoCSP with M3GNet.} Despite inaccurate, ParetoCSP with M3GNet generated structures closer to the corresponding ground truth structures than any other algorithms.}
     \label{fig:failure}
\end{figure}

\subsection{Trajectory study}\label{subsec: traj}

\begin{figure}[ht] 
    \centering
    \begin{minipage}[c]{0.33\textwidth}
        \centering
        \includegraphics[width=\textwidth]{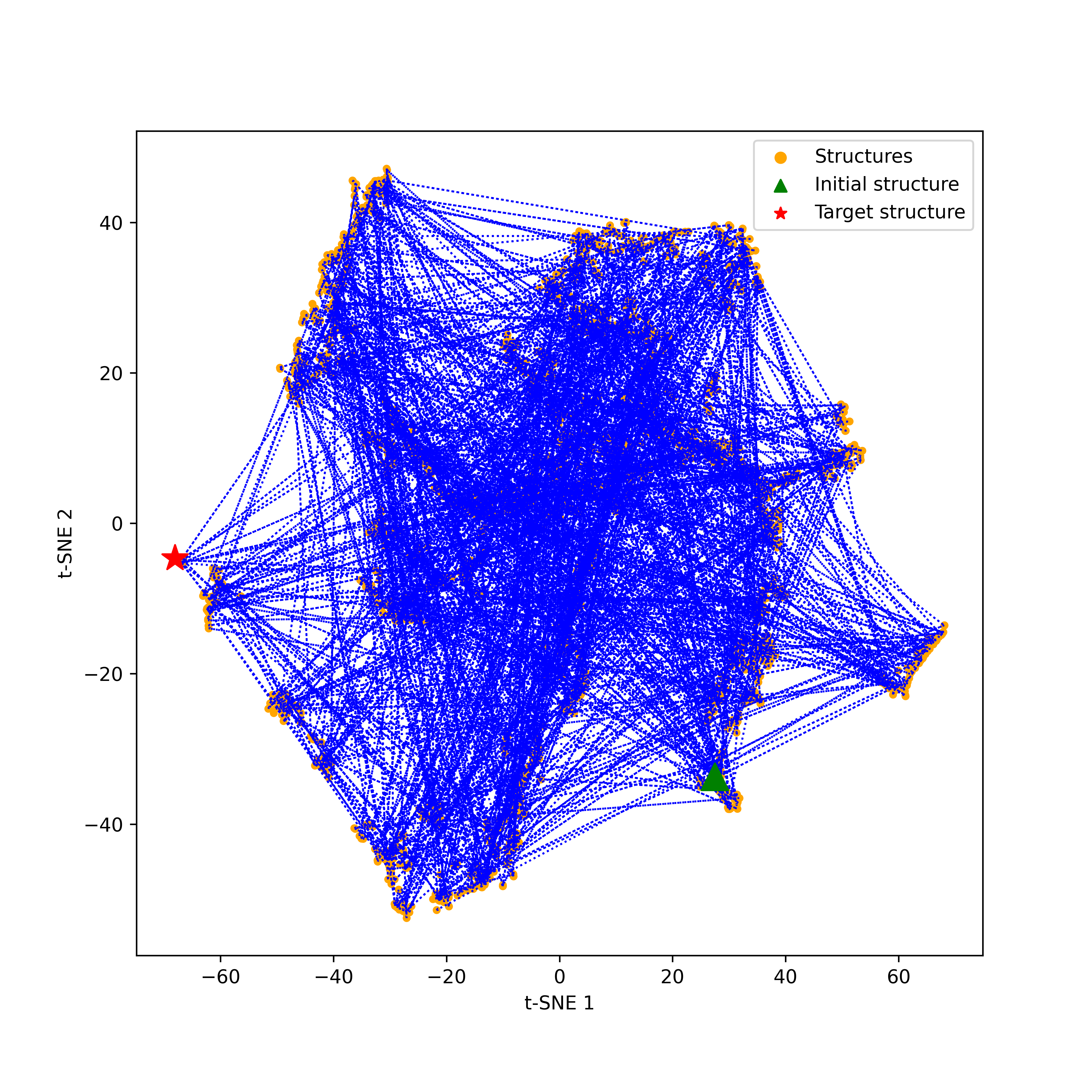}
        \subcaption{SrTiO$_3$ (ParetoCSP).}
        \label{fig: srtio3_step_pc}
    \end{minipage}
    \begin{minipage}[c]{0.33\textwidth}
        \centering
        \includegraphics[width=\textwidth]{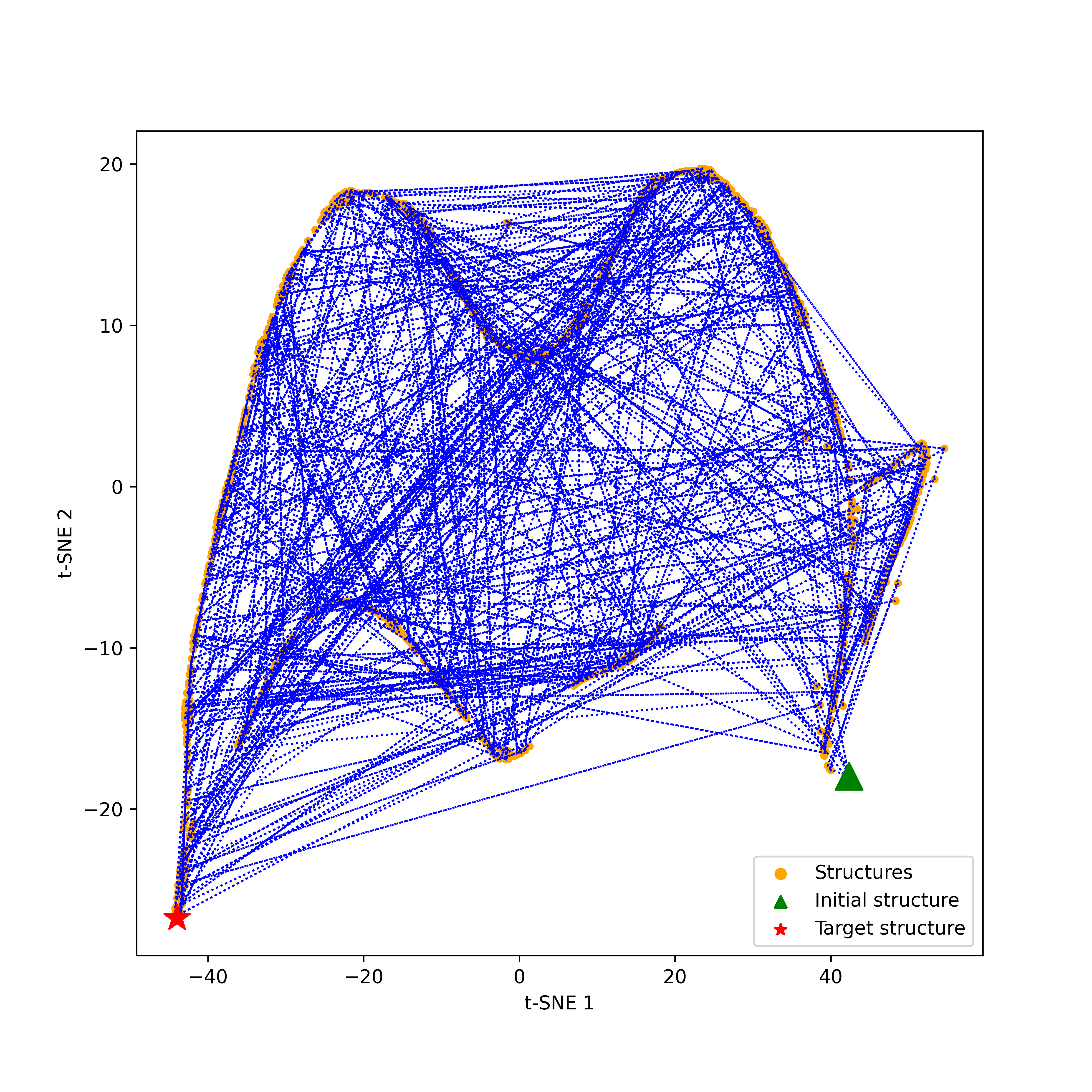}
        \subcaption{SrTiO$_3$ (GN-OA with MEGNet).}
        \label{fig: srtio3_step_bo}
    \end{minipage}
    \begin{minipage}[c]{0.33\textwidth}
        \centering
        \includegraphics[width=\textwidth]{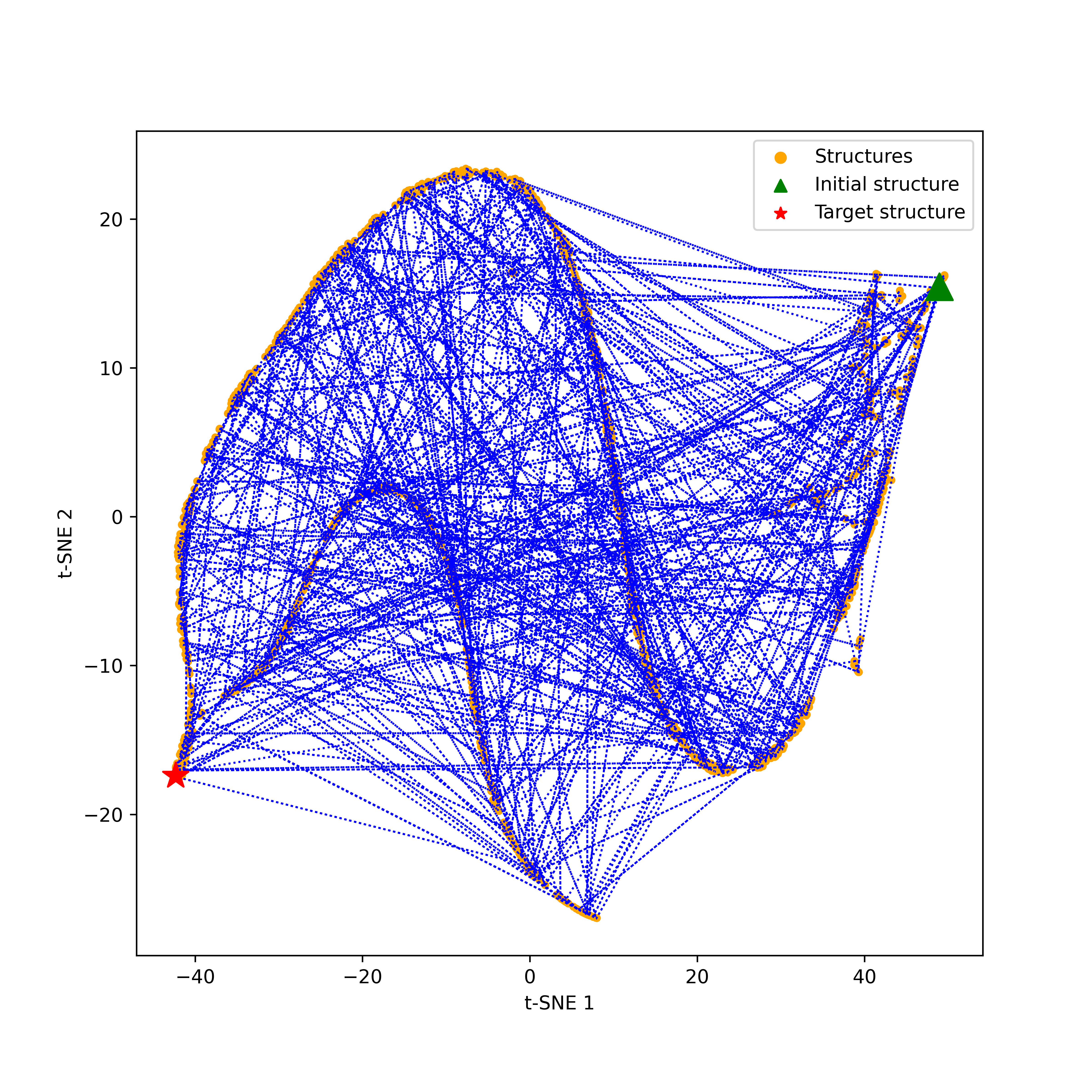}
        \subcaption{SrTiO$_3$ (GN-OA with M3GNet).}
        \label{fig: srtio3_step_bo_m3}
    \end{minipage}\\

    \begin{minipage}[c]{0.33\textwidth}
        \centering
        \includegraphics[width=\textwidth]{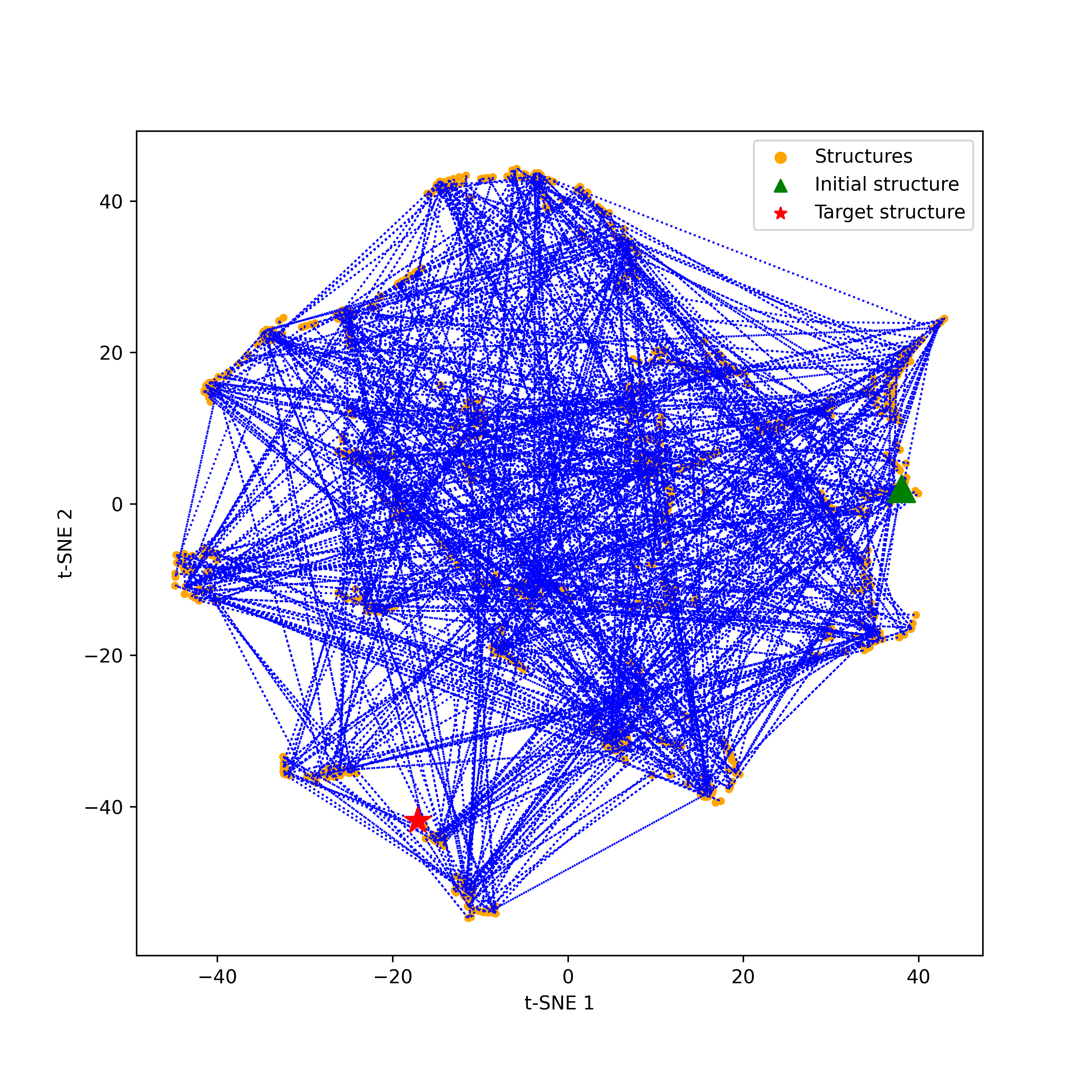}
        \subcaption{MnAlCuPd (ParetoCSP)}
        \label{fig: mnalcupd_step_pc}
    \end{minipage}
    \begin{minipage}[c]{0.33\textwidth}
        \centering
        \includegraphics[width=\textwidth]{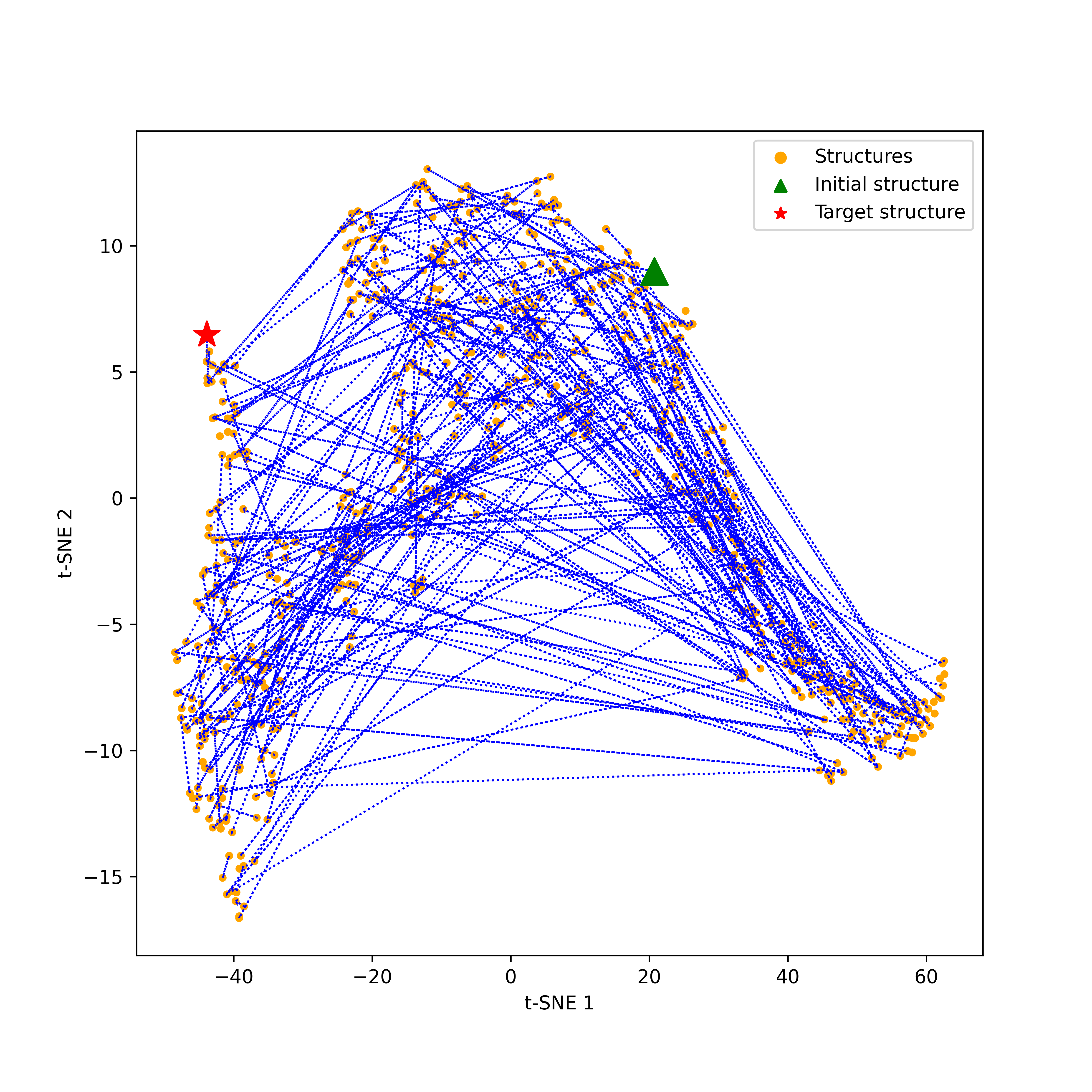}
        \subcaption{MnAlCuPd (GN-OA with MEGNet)}
        \label{fig: mnalcupd_step_bo}
    \end{minipage}
    \begin{minipage}[c]{0.33\textwidth}
        \centering
        \includegraphics[width=\textwidth]{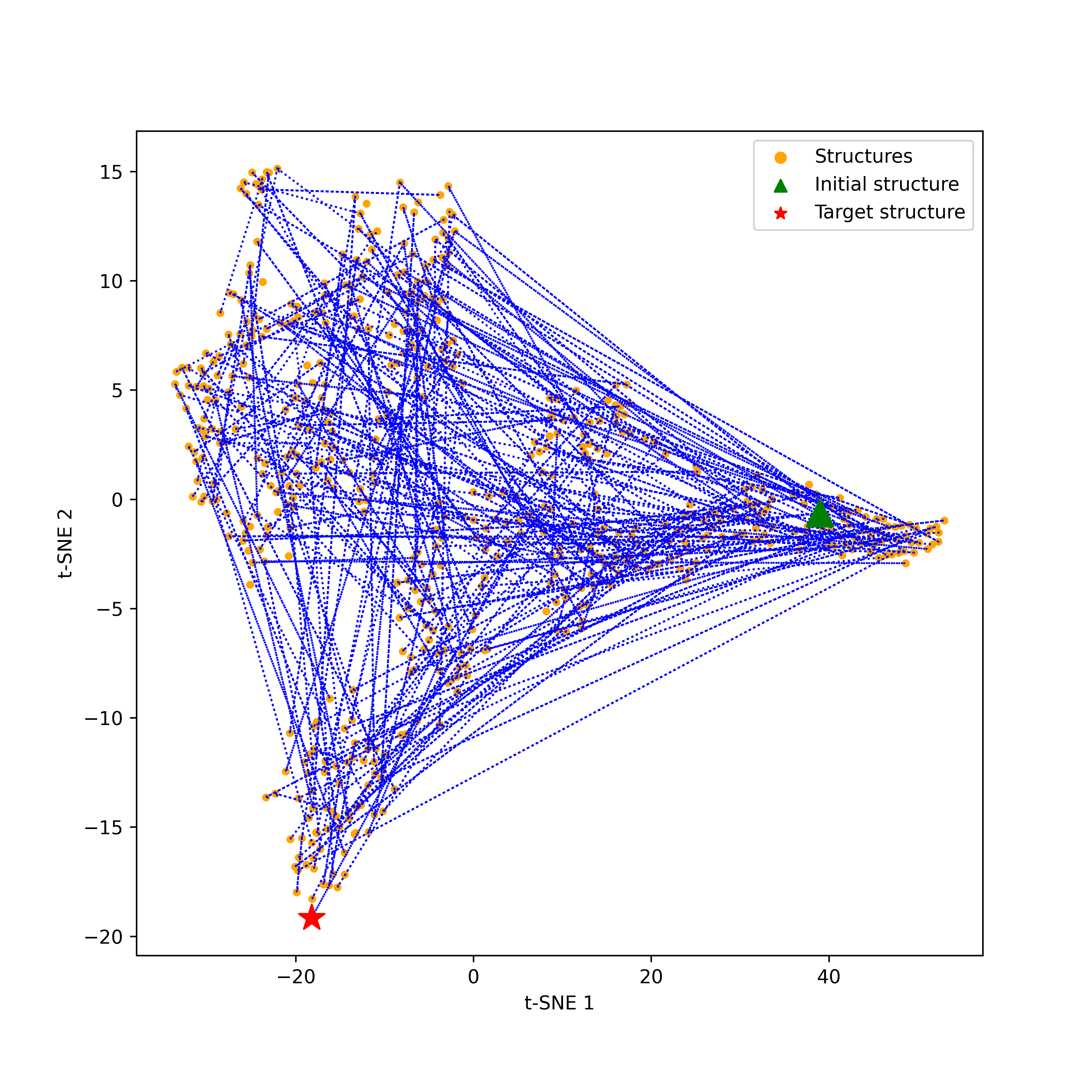}
        \subcaption{MnAlCuPd (GN-OA with M3GNet)}
        \label{fig: mnalcupd_step_bo_m3}
    \end{minipage}\\

    \begin{minipage}[c]{0.49\textwidth}
        \centering
        \includegraphics[width=\textwidth]{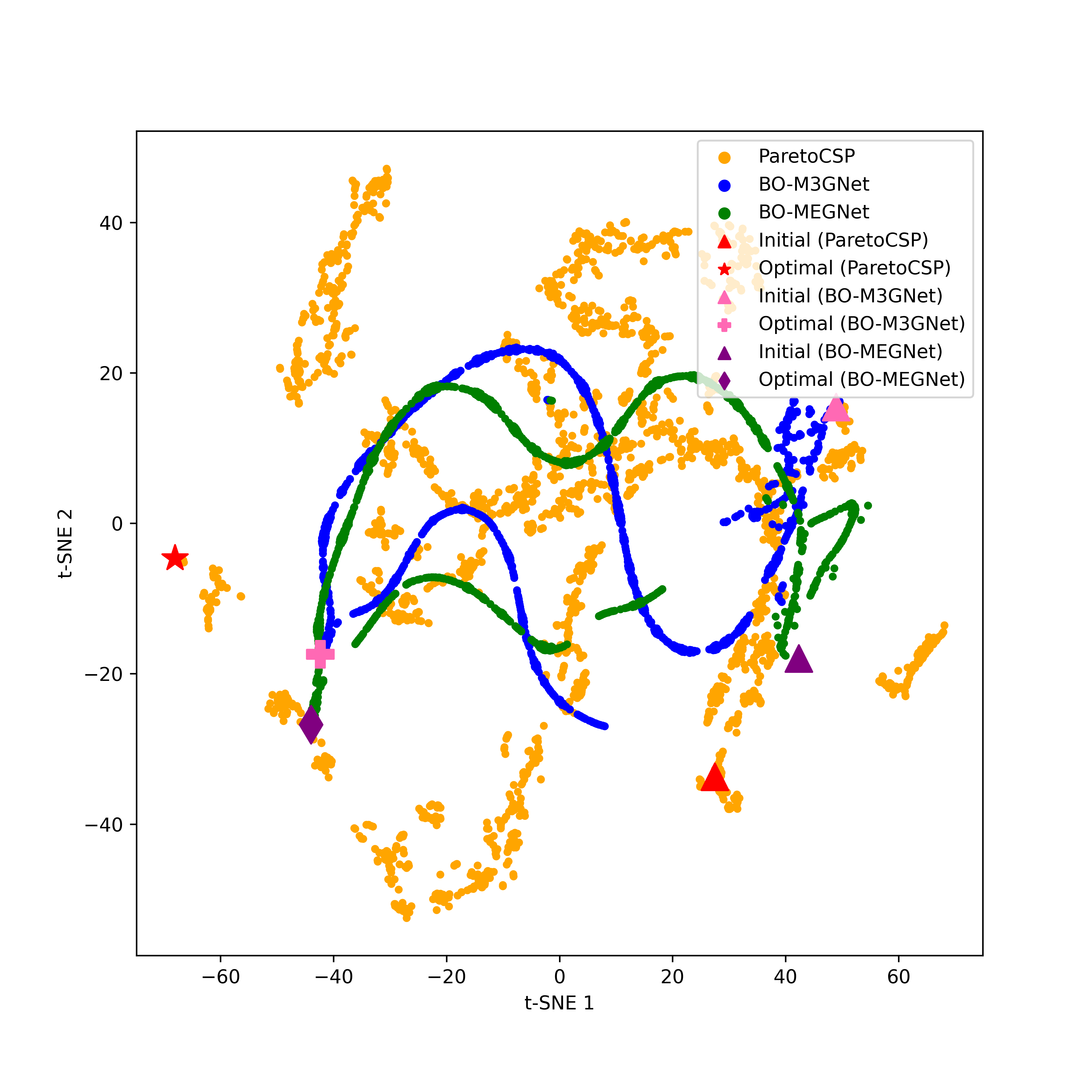}
        \subcaption{SrTiO$_3$ (all)}
        \label{fig: srtio3_step_all}
    \end{minipage}
    \begin{minipage}[c]{0.49\textwidth}
        \centering
        \includegraphics[width=\textwidth]{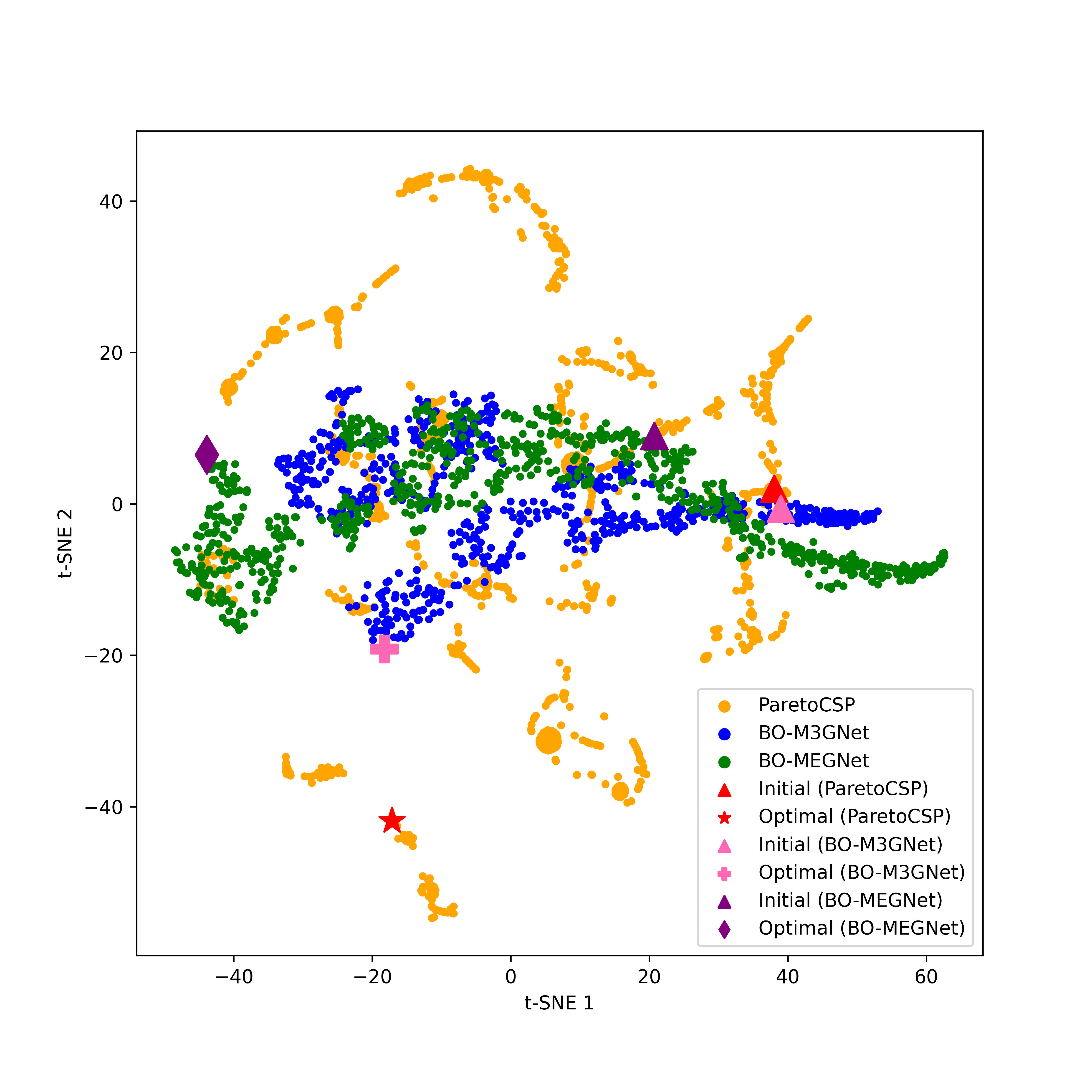}
        \subcaption{MnAlCuPd (all)}
        \label{fig: mnalcupd_step_all}
    \end{minipage}\\

     \caption{\textbf{Trajectories of the traversed structure during search of different CSP algorithms.}\\(a) - (c) shows the trajectory for SrTiO$_3$, and (d) - (f) shows the trajectory for MnAlCuPd. The trajectories were drawn by calculating the distance metrics for the valid structures during the search and mapping them into $2$D space using t-SNE. Two consecutive points were connected if the latter structure had a lower energy than the former one. (g) and (h) show the t-SNE for all three algorithms in the same figure for SrTiO$_3$ and MnAlCuPd, respectively. The initial and optimal structures for all algorithms are marked with different colors and shapes. The points in ParetoCSP's trajectory are more spread out and have more diverse search directions than the other algorithms.}
     \label{fig: trajectory}
\end{figure}

To further understand why ParetoCSP works better than GN-OA algorithm, we utilized the multi-dimensional performance metrics of CSP \cite{cspmetric} to examine the search patterns of both optimization algorithms employed in ParetoCSP and GN-OA. For most of the crystals, the number of valid structures generated by ParetoCSP is enormous. For better visualization, we selected six crystals for this study which had comparatively smaller number of valid structures: SrTiO$_3$, MnAlCuPd, GaNi$_3$, Al$_2$FeCo, Sc$_3$TlC, and SrWNO$_2$. ParetoCSP predicted exact structures of all these crystals, whereas GN-OA failed to predict the structures of MnAlCuPd, Al$_2$FeCo, and SrWNO$_2$. We used a population size of $100$, and total $250$ generations for ParetoCSP. For comparing fairly, we ran a total of $15000$ steps with both GN-OA with MEGNet and M3GNet (GN-OA stopped making progress after $5000$ steps for all of our targets). To analyze the structure search process, we computed the distance metrics between the valid structures and the ground truth structure. These distance features were then mapped into two-dimensional points using t-distributed stochastic neighbor embedding (t-SNE)~\cite{tsne}. The purpose of t-SNE is to map data points from a higher-dimensional space to a lower-dimensional space, typically 2D or 3D, while preserving the pairwise distances between the points. The intuition is that data points that are closer to each other in the higher dimension will remain close to each other after the mapping to the lower dimension. Subsequently, we visualized the trajectories of the structures during the search by connecting consecutive points if the latter structure had a lower energy than the former one. We presented the trajectories for SrTiO$_3$ and MnAlCuPd in Fig.~\ref{fig: trajectory}, and the rest are shown in Supplemental Fig. S3 (see Supplementary Fig. S4 and S5 for trajectory figures without arrows for better visualization of structure mapping). The initial points are represented by green triangles, while the ground truth structures are denoted by red stars. First, the distributions of the generated valid structures over the search by ParetoCSP and GN-OA are very different (Fig.~\ref{fig: srtio3_step_pc} and \ref{fig: mnalcupd_step_pc} versus Fig.\ref{fig: srtio3_step_bo}, \ref{fig: srtio3_step_bo_m3}, \ref{fig: mnalcupd_step_bo}, \ref{fig: mnalcupd_step_bo_m3}). ParetoCSP's distribution are much more diverse while the GN-OA's generated structures tend to be located in a shallow region (Fig.\ref{fig: srtio3_step_all}), indicating that the algorithm can only generate valid structures in a focused path. This is presumably due to the single point search characteristic of the BO algorithm. While a focused search is good when the direction is correct, it runs a high risk of getting trapped in the channeled path and thus loses its structure search capability. These assumptions become more visible from closely looking at Fig.~\ref{fig: srtio3_step_all} and \ref{fig: mnalcupd_step_all} where t-SNE for all three algorithms are drawn in the same figure (see Supplementary Fig. S6 for combined t-SNE for other chosen crystals). We can see that points generated by ParetoCSP are more spread out and have more diverse search directions than other algorithms which ensures its more higher structure search performance. This may explain ParetoCSP's success and GN-OA's failure in predicting structures of MnAlCuPd, Al$_2$FeCo, and SrWNO$_2$.

Another way to understand the structure search efficiency of ParetoCSP and GN-OA is to check the number of valid structures during the search process. 
ParetoCSP generated $2492$, $1518$, $2248$, $2873$, $1843$, and $1633$ valid structures in predicting SrTiO$_3$, MnAlCuPd, GaNi$_3$, Al$_2$FeCo, Sc$_3$TlC, and SrWNO$_2$, respectively, while the original GN-OA with MEGNet generated only $1003$, $681$, $1701$, $1350$, $1499$, and $1066$ valid structures for the same three targets, respectively. GN-OA with M3GNet, instead generated a little bit more valid structures for SrTiO$_3$ (1049), GaNi$_3$ (2044), and Al$_2$FeCo (1475) but fewer for MnAlCuPd ($569$), Sc$_3$TlC ($1165$), and SrWNO$_2$ ($955$). The number of valid structures generated by both GN-OA algorithms are significantly smaller compared to those of our ParetoCSP, indicating that the superiority of ParetoCSP may lie in its capability to make effective search by generating more valid structures. According to the findings of \cite{gnoa}, this showed that our ParetoCSP's AFPO-based GA search function performed much better than BO. Overall, GN-OA struggled to generate valid structures during the search process and wasted a majority of the search dealing with invalid structures. Moreover, the higher percentage of valid structures generated and more diverse search function of ParetoCSP may have contributed to its higher probability of finding the exact structures.

\FloatBarrier

\section{Discussion}
We present ParetoCSP, a CSP algorithm which combines an AFPO enhanced multi-objective GA as an effective structure search function and M3GNet universal IAP as a constructive final energy predictor to achieve efficient structure search for CSP. The objective is to effectively capture the complex relationships between atomic configurations and their corresponding energies. Firstly, ParetoCSP uses the age of a population as a separate optimization criterion. This leads the algorithm to treat the age as a separate dimension in the multi-objective Pareto front where the GA aims to generate structures to minimize the final energy per atom, as well as having low genotypic age. According to the finding of \cite{afpo}, this provides a more extensive search process which enables the NSGA-III to perform better as shown in the trajectory results in Section~\ref{subsec: traj}, where we see that ParetoCSP generated a lot more valid structures during the search process than other evaluated CSP algorithms. This demonstrates the effective exploration of the crystal structure space by ParetoCSP and efficient identification of the most stable structures. 

Overall, we found that ParetoCSP remarkably outperforms the GN-OA algorithm by a factor of $2.562$ and overall achieved $74.55\%$ accuracy. The comprehensive experimentation was carried out on $55$ benchmark sets consisting of diverse space groups, which shows that the algorithm can efficiently handle a wide range of crystal systems, including complex ternary and quarternary compounds, whereas GN-OA performed poorly on the quarternary crystals, and most of the ternary crystals. Moreover, a majority of them belongs to the cubic crystals system, proving GN-OA's lack of capability of explore the structure space of diverse crystal systems. However, all the algorithms show poor performance for crystals belonging to the orthorhombic and monoclinic crystal systems. This performance limits of ParetoCSP can be attributed to either the optimization algorithm or the ML potential.

First we found that for both ParetoCSP and GN-OA, the search process tends to generate a majority of invalid structures even though ParetoCSP works much better than GN-OA. These invalid structures are a waste of search time. Better algorithms that consider crystal symmetry or data-driven generative models may be developed to improve the percentage of valid structures and increase the search efficiency during the search process. 
In ParetoCSP, the M3GNet IAP is used as the final energy predictor during the search process and structure relaxer after finishing the search process. Compared to MEGNet, M3GNet IAP is proven to be a better choice since after replacing GN-OA's MEGNet with M3GNet IAP, its performance can be improved by a factor of $1.5$. Overall, our results suggest the importance of developing stronger universal ML potentials in modern CSP algorithm development.  Other IAP models such as TeaNet~\cite{teanet} can be experimented to check whether better performance can be achieved with ParetoCSP and can be compared to the results with M3GNet. Unlike GN-OA, ParetoCSP performs a further refinement of the output structure which helped generate exact structures. We used M3GNet IAP for the structure relaxation. More advanced structure relaxation methods can be tested instead to get better performance.

For the first time, we have used a set of seven quantitative performance metrics to compare and investigate algorithm performances of ParetoCSP and the baselines. We can see from Table~\ref{table:result_m3gnet} that each of the unsuccessful predictions had at least one of the performance metrics value larger than the ground truth value. Additionally, Fig.~\ref{fig:success} shows that ParetoCSP with M3GNet generated better solutions than any other baseline CSP algorithms as they had much lower performance metric distances (errors) than others. Furthermore, the performance metrics also show that even though ParetoCSP was unable to predict $14$ crystal structures, it still produced better quality structures compared to other CSP algorithms. They can also be used to show for a specific crystal whether the algorithm is on the right track to predict its structure or not.

Inspired by the great success of AlphaFold2~\cite{protein2} for protein structure prediction, which does not rely first principles calculations, we believe that data-driven CSP algorithms based on ML deep neural network energy models have big potential and can reach the same level as AlphaFold2. For this reason, we have focused on the performance comparison with the state-of-the-art GN-OA, a ML potential based CSP algorithm and we did not compare our results with CALYPSO~\cite{calypso} and USPEX~\cite{uspex}, despite that USPEX also utilizes evolutionary algorithms like ours. These algorithms are extremely slow and are not scalable to complex crystals as they depend on ab-initio energy calculations, which is computationally very expensive and slow. Currently, they can only deal with simple chemical systems or relatively small crystals ($< 10$ atoms in the unit cell) which is a major disadvantage.

\section{Conclusion}

We have introduced an innovative CSP algorithm named ParetoCSP, which synergizes two key components: the multi-objective GA employing age-fitness Pareto optimization and the M3GNet IAP, for predicting the most stable crystalline material structures. The AFPO-based GA effectively functions as a structure search algorithm, complemented by the M3GNet IAP's role as an efficient final energy predictor that guides the search process. Through comprehensive experimentation involving $55$ benchmark crystals, our algorithm's potency has been demonstrated, notably surpassing GN-OA with MEGNet and GN-OA with M3GNet by substantial factors of $2.562$ and $1.71$, respectively. Utilizing benchmark performance metrics, we have provided an in-depth analysis of the quality of structures generated by our algorithm. Furthermore, we have quantitatively depicted deviations from the ground truth structure for failure cases across all algorithms, highlighting ParetoCSP's superior performance in this aspect as well. By means of a trajectory analysis of the generated structures, we have established that ParetoCSP produces a greater percentage of valid structures compared to GN-OA during the search process due to its enhanced search algorithm. Given these significant progress, we believe that ML potential based CSP algorithms such as ParetoCSP hold immense promise for advancing CSP's boundaries and facilitating the discovery of novel materials with desired properties.

\section*{Contribution}
Conceptualization, J.H.; methodology,S.O., J.H., L.W.; software,S.O., J.H. ; resources, J.H.; writing--original draft preparation, S.O., J.H., L.W.; writing--review and editing, J.H and L.W.; visualization, S.O. ; supervision, J.H.;  funding acquisition, J.H.

\section*{Acknowledgement}
The research reported in this work was supported in part by National Science Foundation under the grant 10013216 and 2311202. The views, perspectives, and content do not necessarily represent the official views of the NSF.

\bibliographystyle{ieeetr}
\bibliography{references}

\end{document}